\documentclass[11pt,a4paper]{article}
\pdfoutput=1
\usepackage{jheppub_kim}

\usepackage{multirow, graphicx,amssymb,url,mathrsfs,amsmath}
\usepackage{wrapfig,boxedminipage,setspace,subfigure,epsfig}
\usepackage{amsxtra,amstext,latexsym,dsfont,amsfonts}
\usepackage{color,eucal}
\usepackage[dvipsnames]{xcolor}
\usepackage{float}
\usepackage{slashed}

\usepackage{tabularx, array}
\newcolumntype{L}[1]{>{\raggedright\arraybackslash}p{#1}}
\newcolumntype{C}[1]{>{\centering\arraybackslash}p{#1}}
\newcolumntype{R}[1]{>{\raggedleft\arraybackslash}p{#1}}

       \def\G  {\Gamma}

\renewcommand{\d}{\delta}

\newcommand{\dd}{\mathrm{d}}

\newcommand{\grad}{\nabla}

\newcommand{\abs}[1]{\left| #1 \right|}

\newcommand{\tb}{\mbox{${\tilde{\beta}}$}}
\newcommand{\bb}{\mbox{${\bar{\beta}}$}}
\newcommand{\bT}{\mbox{${\bar{T}}$}}

\newcommand{\tT}{\mbox{$\tilde{T}$}}

\newcommand{\tmu}{{\tilde{\mu}}}

\newcommand{\udt}{{\underline{t}}}
\newcommand{\udr}{{\underline{r}}}
\newcommand{\udx}{{\underline{x}}}
\newcommand{\udy}{{\underline{y}}}

\title{Holographic Spectral Functions with Momentum Relaxation}

\author[a]{Hyun-Sik Jeong,}
\author[a]{Keun-Young Kim,}
\author[b]{Yunseok Seo,}
\author[c]{Sang-Jin Sin}
\author[d]{and Shang-Yu Wu}

\emailAdd{hyunsik@gist.ac.kr}
\emailAdd{fortoe@gist.ac.kr}
\emailAdd{yseo@gist.ac.kr}
\emailAdd{sjsin@hanyang.ac.kr}
\emailAdd{loganwu@gmail.com}

\affiliation[a]{ School of Physics and Chemistry, Gwangju Institute of Science and Technology, \\
123 Cheomdan-gwagiro, Gwangju 61005, Korea}
\affiliation[b]{GIST College, Gwangju Institute of Science and Technology, Gwangju 61005, Korea}
\affiliation[c]{Department of Physics, Hanyang University, Seoul 133-791, Korea}
\affiliation[d]{Department of Physics, Chung-Yuan Christian University, Chung Li District, Taoyuan City 32023, Taiwan}

\abstract{
We study (fermionic) spectral functions in two holographic models, the Gubser-Rocha-linear axion model and the linear axion model, where translational symmetry is broken by axion fields linear to the boundary coordinates ($\psi_{I}=\beta \delta_{Ii} x^{i}$). Here, $\beta$ corresponds to the strength of momentum relaxation. The spectral function is computed by the fermionic Green's function of the bulk Dirac equation, where a fermion mass, $m$, and a dipole coupling, $p$, are introduced as input parameters. By classifying the shape of spectral functions, we construct complete phase diagrams in ($m,p,\beta$) space for both models. 
We find that two phase diagrams are similar even though their background geometries are different. We also find that the effect of momentum relaxation on the (spectral function) phases of two models are similar even though the effect of momentum relaxation on the DC conductivities of two models are very different. We suspect that this is because holographic fermion does not back-react to geometry in our framework. 

}

\begin{document}

\maketitle

\section{Introduction}
One of the important milestones in condensed matter physics is Landau's Fermi liquid theory because it provides us a way to understand almost all metals, semi-conductors, superconductors and superfluids. Since 1980's some exotic materials have been found, which cannot be explained by the Fermi liquid theory, called non-Fermi liquid. In addition to non-Fermi liquid, various phases of matter are discovered in strongly correlated systems such as the strange metals and high temperature superconductors~\cite{zaanen:2006aana, Basov:2011aa, Iqbal:2011ae}. 
However, theoretical frameworks to describe them are still not complete.

The gauge/gravity duality (or AdS/CFT correspondence, or holographic methods) have provided effective tools to study such exotic phases in strongly correlated materials~\cite{Zaanen:2015oix, Hartnoll:2009sz, Herzog:2009xv}, which states that the strongly correlated condensed matter physics can be mapped to the classical gravity physics.
Using this method, there are many studies of fermionic response in strongly correlated system. We refer the reader to \cite{Lee:2008xf, Liu:2009dm, Cubrovic:2009ye, Faulkner:2009wj} for pioneering works and \cite{Iqbal:2011ae} for a comprehensive review. In particular, studying holographic ``spectral function'' could be an important test for the application of gauge/gravity duality because it can be compared with measurements of Angle Resolved Photoemission Spectroscopy (ARPES) or Scanning Tunneling Microscopy (STM).

\paragraph{Momentum relaxation:}
As well as spectral function, conductivity is one of the most important and widely studied experimental observables. To investigate conductivity in holography, it is important to consider holographic models with translational symmetry broken, because, with translation symmetry, momentum cannot be relaxed at finite density and conductivity is simply infinite. 

Depending on the method to break translational symmetry, holographic models can be divided into two classes: homogeneous and inhomogeneous models. ``Homogeneous'' means the spacetime geometry is only functions of the holographic direction and independent of field theory directions. Otherwise, models are called inhomogeneous. 

For example, homogeneous models include i) helical lattice~\cite{Donos:2012js, Donos:2014oha, Donos:2014gya}, ii) Q-lattice~\cite{Donos:2013eha, Donos:2014uba}, iii) linear axion model~\cite{Andrade:2013gsa, Gouteraux:2014hca, Taylor:2014tka, Kim:2014bza, Kim:2015wba}. In particular, in the linear axion model, translational symmetry is broken by massless scalar fields linear in spatial coordinates. This type of models is related to the St\"uckelburg formulation of a massive gravity theory~\cite{Vegh:2013sk, Davison:2013jba, Blake:2013bqa, Blake:2013owa}.
Inhomogeneous models are obtained by imposing periodic boundary conditions on a scalar field (scalar lattice) or chemical potential (ionic lattice). See, for pioneering works, ~\cite{Horowitz:2012ky, Horowitz:2012gs}.
The models can be also classified by symmetry breaking mechanism: i) explicit breaking, ii) spontaneous symmetry breaking for pair density wave, charge density wave, spin density wave phases.\footnote{These are representative examples of striped quantum phases in strongly correlated electron systems~\cite{Fradkin_2015} and they are believed to play a crucial role in strongly correlated systems, in particular, in high temperature superconductors. } 
For example, for spontaneously generated inhomogeneous lattice model see~\cite{Cremonini:2016rbd, Cremonini:2017usb}.

In inhomogeneous models, because of the periodic boundary conditions of fields, the equations become partial differential equations (PDE), while, in homogeneous models, the equations are ordinary differential equations (ODE). Thus, homogeneous models are technically more tractable than inhomogeneous models, and sometimes allow partly analytic treatment and intuitions. 
In this paper, we focus on the homogeneous models based on linear axion models.

\paragraph{Spectral function with momentum relaxation:}
As we mentioned, broken translation symmetry is now an essential ingredient in holographic models, to give momentum relaxation and the finite conductivity. 
Therefore, for consistency, it is important to study spectral function in holographic models with translational symmetry broken. To the best of our knowledge, we collect all relevant literatures: \cite{Liu:2012tr, Ling:2013aya, Ling:2014bda, Bagrov:2016cnr, Fang:2016wqe, Cremonini:2018xgj, Cremonini:2019fzz, Balm:2019dxk}.
For the comparison of models, see Table \ref{SUM}, where there are three parameters: fermion mass ($m$), dipole coupling ($p$), strength of momentum relaxation ($\beta$). The precise meaning of $m$ and $p$ will be given in \eqref{SpinorAction}. The precise meanings of $\beta$ are different model by model, but the main point is it represents  the strength of momentum relaxation.

Let us categorize the models in Table  \ref{SUM} into two classes: homogeneous and inhomogeneous holographic models.
First, \cite{Ling:2014bda, Fang:2016wqe, Bagrov:2016cnr} deal with homogeneous holographic models. It is known that the dipole coupling can open a Mott-like gap~\cite{Edalati:2010ge,Edalati:2010ww,Vanacore:2015poa}. The main theme of \cite{Ling:2014bda, Fang:2016wqe} is  the effect of momentum relaxation on the gap-generation mechanism of the dipole coupling.  
\cite{Ling:2014bda} found that the insulating phase opens the Mott gap much easier than metallic phase in the Q-lattice model and \cite{Fang:2016wqe} reports that a bigger dipole coupling is required to open a gap at larger momentum relaxation in the massive gravity model. \cite{Bagrov:2016cnr} found that the spectral functions are suppressed at large momentum relaxation in helical lattice model, where there is no dipole coupling. 

Next, \cite{Liu:2012tr, Ling:2013aya, Cremonini:2018xgj, Cremonini:2019fzz, Balm:2019dxk} deal with inhomogeneous holographic models. Their main interest is the shape-change of the Fermi surface when there is momentum relaxation (lattice effect) along one direction, so it is kind of anisotropic effect. 
\cite{Liu:2012tr, Ling:2013aya} found that the shape of the Fermi surface is changed from circle to ellipse at weak lattice potential, and \cite{Cremonini:2018xgj, Cremonini:2019fzz, Balm:2019dxk} discovered a disappearance of Fermi surface along the symmetry breaking direction when the lattice strength increases.

\begin{table}[]
\begin{tabular}{| >{\centering\arraybackslash}m{1.7cm} | >{\centering\arraybackslash}m{5.2cm} | >{\centering\arraybackslash}m{3.1cm} | >{\centering\arraybackslash}m{1.4cm} | >{\centering\arraybackslash}m{1.4cm} |}
\hline
 References  &  Models   & $m$   & $p$  & $ \beta$\\
 \hline
 \hline
  \cite{Bagrov:2016cnr}     &\hspace{0.1cm} Helical lattice   &$m=0 \,, \quad m=2.5$    & $p=0$ &  $\beta \neq 0$\\ 
 \hline
 \cite{Ling:2014bda}        &\hspace{0.1cm} Q-lattice   & $m=0$    & $p\neq0$  &  $\beta \neq 0$\\ 
 \hline
\cite{Fang:2016wqe}         & Non-linear massive gravity        & $m=0$    & $p\neq0$  & $\beta \neq 0$\\ 
 \hline
 \hline 
  \cite{Liu:2012tr, Ling:2013aya, Cremonini:2018xgj, Cremonini:2019fzz}                   &  Spontaneously or Explicit generated lattice  &  $m=0$    &  $p=0$   &   $\beta \neq 0$ \\
 \hline
\cite{Balm:2019dxk}         &\hspace{0.1cm} Ionic lattice              &$\small{m= 1/4, m=3/4}$    & $p=0$ &  $\beta \neq 0$\\ 
 \hline
\end{tabular}
\caption{Studies of holographic fermions \textit{with} momentum relaxation}
\label{SUM}
\end{table}
\begin{table}[]
\begin{tabular}{| C{1.7cm} | C{5.2cm} | C{3.1cm} | C{1.4cm} | C{1.4cm} |}
\hline
 References  & Models   & $m$   & $p$  & $\beta$\\ 
 \hline
 \hline
    \cite{Seo:2018hrc}                      & RN-AdS   &$m\neq0$    & $p\neq0$   & $\beta = 0$\\
 \hline
     \cite{Wu:2011cy}    & Gubser-Rocha model   &$m=0$    & $p=0$   & $\beta = 0$\\
 \hline
     \cite{Li:2011sh, Wen:2012ur}    & Gubser-Rocha model   &$m=0$    & $p\neq0$   & $\beta = 0$\\
 \hline
\end{tabular}
\caption{Studies of holographic fermions \textit{without} momentum relaxation}
\label{SUM2}
\end{table}

\paragraph{Motivations of this paper:} One of our motivations is to  investigate the effect of momentum relaxation on spectral functions more systematically and thoroughly in two senses. 

First, as shown in Table. \ref{SUM} the effect of momentum relaxation has been investigated in part of parameter space $(m,p,\beta)$: i) most works focus on $m=0$ case or a few finite values of $m$; ii) for the cases with $p \ne 0$ and/or $\beta \ne 0$, only some values are chosen for illustrative purposes; iii) the case with $(m\ne0, \, p\ne0)$ has not been considered yet. Therefore, we will explore whole parameter range in $(m, p, \beta)$ space to understand momentum relaxation effect from weak ($\beta \ll 1$) to strong ($\beta \gg 1$) case for various values of $(m, p)$.\footnote{In \cite{Bagrov:2016cnr, Ling:2014bda, Cremonini:2018xgj, Cremonini:2019fzz}, it was reported that the spectral functions are suppressed if there is momentum relaxation in limited range of parameters ($m,p,\beta$).}

Second, recently it has been shown that, in the Reissner-Nordstrom AdS (RN-AdS) black holes geometry, there are various phases in $(m, p)$ space in terms of spectral density~\cite{Seo:2018hrc} in addition to the well known non-Fermi liquid or gapless phase~\cite{Liu:2009dm,Faulkner:2009wj,Faulkner:2011tm,Faulkner:2013bna} and gapped phase~\cite{Edalati:2010ge,Edalati:2010ww,Vanacore:2015poa,Seo:2018hrc}. New phases found in \cite{Seo:2018hrc} include various types of gapless phase, pseudogap phase and gapped phase.
However, it has not been studied yet how these phases are affected by momentum relaxation. 
These phases can be enhanced/suppressed or new phases may arise in principle.  To investigate this problem we will study the RN-AdS model with linear-axion to impose momentum relaxation. This model is called ``linear axion model'' {in short~\cite{Andrade:2013gsa}.}

Another motivation of our work is to investigate if there is any relation between the background geometry and spectral function, a ``probe'' of the background geometry. 
Note that all works presented in Table \ref{SUM} are based on the RN-AdS geometry in the sense the geometries become RN-AdS if $\beta \rightarrow 0$. Let us call them ``RN-AdS-like''. 
Thus, it is important to study the phase diagram for spectral function in another background different from RN-AdS-like models. A good candidate is a Gubser-Rocha model~\cite{Gubser:2009qt} with linear axion~\cite{Davison:2013txa, Kim:2017dgz}, which we will call ``Gubser-Rocha-linear axion'' model. The Gubser-Rocha-linear axion model may be phenomenologically more appealing than RN-AdS-like models because it has zero entropy at zero temperature~\cite{Gubser:2009qt, Davison:2013txa} and also exhibits linear-$T$-resistivity (an interesting universal property of strange metals\footnote{See \cite{Ahn:2019lrh} for holographic models showing linear-$T$ resistivity up to high temperature.})~\cite{Davison:2013txa, Jeong:2018tua} contrary to RN-AdS-like models.

Note that there is no work on the spectral function in the Gubser-Rocha-linear axion model even though there is only a few works~\cite{Li:2011sh, Wu:2011cy, Wen:2012ur}  in the Gubser-Rocha model (no linear axion) only for a small range in ($m,p$) where there is no momentum relaxation. See Table \ref{SUM2}.  In order to make a complete comparison between the RN-AdS-like model and the Gubser-Rocha-linear axion model and to see if there is any relation between the spectral function and background geometry, we will study the spectral functions in the Gubser-Rocha-linear axion model in full ($m,p,\beta$) space. 

In summary, in this paper, we study two simple holographic models with momentum relaxation: the linear axion model~\cite{Andrade:2013gsa} and the Gubser-Rocha-linear axion model~\cite{Davison:2013txa, Kim:2017dgz}. Both models contain axion field which breaks the translational symmetry so renders the conductivity finite. Both have an advantage of allowing the analytic solution.

This paper is organized as follows. 
In section 2, we review the methods to compute the holographic spectral function.
In section 3, we compute the spectral function in the Gubser-Rocha-linear axion model. 
In section 4, we study the spectral function in the linear axion model and compare it with the Gubser-Rocha-linear axion model. 
In section 5, we conclude.

\section{Method} \label{method123}
\subsection{Dirac equation and Green's function}
In this section, we review the fundamentals of the fermionic spectral function in a holographic framework ~\cite{Liu:2009dm, Gubser:2009dt, Seo:2018hrc}.
We start with a fermion probe bulk action with the fermion mass $m$ and the dipole interaction $p$
\begin{equation} \label{SpinorAction}
\begin{split}
S_{\text{spinor}} &= \int \dd^4 x \sqrt{-g} \, i  \bar{\psi} \left( \Gamma^{M} D_{M} -m - \, \frac{i  p}{2}\Gamma^{MN} F_{MN}  \right)\psi  \,, \\
\Gamma^{MN} &= \frac{1}{2} [\Gamma^{M},\Gamma^{N}] \,,
\end{split}
\end{equation}
where $\Gamma^{M}$  and $D_{M}$ is the Gamma matrices \eqref{GAMMAMA} and the covariant derivative in a curved spacetime \eqref{COV} respectively.
We consider this probe fermion in the background metric and gauge field
\begin{equation} \label{Generalmetric}
\begin{split}
\dd s^2 &= -g_{tt}(r) \dd t^2 + g_{rr}(r) \dd r^2 + g_{xx}(r) \dd x^2  + g_{yy}(r) \dd y^2  \,, \\
         A &= A_{t}(r) \dd t \,,
\end{split}
\end{equation}
where $A$ is a U(1) gauge field with the field strength $F = \dd A$.

The Gamma matrices can be expressed in terms of the tangent space indices as follows:
\begin{equation} \label{GAMMAMA}
\begin{split}
\Gamma^{M} = \Gamma^{a}e_{a}^{M} \,, \quad \Gamma^{MN} = \Gamma^{ab}e_{a}^{M}e_{b}^{N} \,,
\end{split}
\end{equation}
where $e_{a}^{M}$ is the inverse veinbein satisfying $g_{MN}(x)=\eta_{ab} \, e_{M}^{a}(x)e_{M}^{b}(x)$ with the flat spacetime metric $\eta_{ab}=(-1,1,1,1)$. Here, $M,N$ and $a,b$ indicate the bulk spacetime and the tangent spacetime respectively. We will also use underline indices to refer to tangent spacetime. In other words, $M,N$=($t,r,x,y$) and $a,b$=($\udt,\udr,\udx,\udy$).
Note that this Gamma matrices satisfy the Clifford algebra, $ \{\Gamma^{a},\Gamma^{b}\}=2\eta^{ab}$.
The Dirac adjoint ($\bar{\psi}$) and the covariant derivative ($D_{M}$) in \eqref{SpinorAction} are given as
\begin{equation} \label{COV}
\begin{split}
\bar{\psi} = \psi^{\dagger} \Gamma^{\udt} \,,\quad D_{M} = \partial_{M} + \frac{1}{4}\omega_{abM} \G^{ab}-iqA_M \,,
\end{split}
\end{equation}
%
with the spin connection $\omega_{abM}$ determined by the torsion-free condition $\dd e^{a} + \omega_{b}^{a}\wedge e^{b}=0$.\footnote{$\omega_{b}^{a}$ is a shorthand notation for $\omega_{b \, M}^{a} \dd x^{M}$.}
In what follows, we will choose the following veilbein
\begin{equation} \label{veil}
\begin{split}
e_\udt^t = \frac{1}{\sqrt{g_{tt}}} \,,\quad e_\udr^r = \frac{1}{\sqrt{g_{rr}}} \,,\quad e_\udx^x = \frac{1}{\sqrt{g_{xx}}} \,,\quad e_\udy^y = \frac{1}{\sqrt{g_{yy}}} \,,
\end{split}
\end{equation}
and the non-zero components of the spin connection from the torsion-free condition are
\begin{equation} \label{spinconnec}
\begin{split}
\omega_{\udt \, \udr \,t} = -\frac{\partial_{r}\sqrt{g_{tt}}}{\sqrt{g_{rr}}} \,, \quad \omega_{\udx \, \udr \,x} = \frac{\partial_{r}\sqrt{g_{xx}}}{\sqrt{g_{rr}}} \,, \quad \omega_{\udy \, \udr \,y} = \frac{\partial_{r}\sqrt{g_{yy}}}{\sqrt{g_{rr}}} \,.
\end{split}
\end{equation}

With the Gamma matrices in \eqref{GAMMAMA} and the covariant derivative in \eqref{COV},
the bulk Dirac equation \eqref{SpinorAction} reads
\begin{equation} \label{DiracEq1}
\begin{split}
\left( \Gamma^{a}e_{a}^{M} (\partial_{M} + \frac{1}{4}\omega_{ab \, M} \G^{ab}-iqA_M) -m - \, \frac{i  p}{2}\Gamma^{ab}e_{a}^{M}e_{b}^{N} F_{MN}  \right)\psi  = 0 \,.
\end{split}
\end{equation}
After putting the veilbein \eqref{veil} and a spin connection \eqref{spinconnec} into the bulk Dirac equation \eqref{DiracEq1}, we have the equation with the ansatz\footnote{{This ansatz makes the spin connection removed.}} $\psi = (-g\,g^{rr})^{-\frac{1}{4}} \varphi$ as
\begin{equation} \label{DiracEq2}
\begin{split}
\frac{\Gamma^{\udr} \, \partial_{r}\varphi}{\sqrt{g_{rr}}} + \frac{\Gamma^{\udt}\left(\partial_{t} - i q A_{t} \right)\varphi}{\sqrt{g_{tt}}} + \left(\frac{\Gamma^{\udx}\partial_{x}}{\sqrt{g_{xx}}} + \frac{\Gamma^{\udy}\partial_{y}}{\sqrt{g_{yy}}}\right)\varphi - m \, \varphi -\frac{ip \, \Gamma^{\udr\,\udt} A_{t}' \, \varphi}{\sqrt{g_{tt}g_{rr}}} = 0 \,.
\end{split}
\end{equation}
Since we will consider the background metric \eqref{Generalmetric} which has the rotational symmetry in the spatial directions, we can set $(k_{x}, k_{y})=(k, 0)$ without loss of generality.
Therefore, by introducing $\varphi = e^{i(kx-\omega t)} \phi$, we can rewrite the bulk Dirac equation \eqref{DiracEq2} in the Fourier space:
\begin{equation} \label{DiracEq3}
\begin{split}
\sqrt{\frac{g_{xx}}{g_{rr}}}\left( \Gamma^{\udr}\partial_{r} - m\sqrt{g_{rr}} \right)\phi + \left(-i u \Gamma^{\udt} + i k \Gamma^{\udx} \right)\phi - ip \sqrt{\frac{g_{xx}}{g_{tt}\,g_{rr}}}\Gamma^{\udr\,\udt} A_{t}' \phi = 0 \,,
\end{split}
\end{equation}
where
\begin{equation} \label{}
\begin{split}
u = \sqrt{\frac{g_{xx}}{g_{tt}}}(\omega + q A_{t}) \,.
\end{split}
\end{equation}

To study the bulk Dirac equation, it is convenient to use $\phi = \left(\begin{array}{c}\phi_{+} \\\phi_{-}\end{array}\right)$
with the following Gamma matrices~\cite{Liu:2009dm,Gubser:2009dt}:
\begin{equation} \label{GammaMat}
\begin{split}
\Gamma^{\udr} = \left(\begin{array}{cc}1 & 0 \\0 & -1\end{array}\right) \,, \quad \Gamma^{\udt} = \left(\begin{array}{cc}0 & i\sigma_{2} \\i\sigma_{2} & 0\end{array}\right) \,, \quad \Gamma^{\udx} = \left(\begin{array}{cc}0 & \sigma_{1} \\\sigma_{1} & 0\end{array}\right) \,, \quad \Gamma^{\udy} = \left(\begin{array}{cc}0 & \sigma_{3} \\\sigma_{3} & 0\end{array}\right) \,,
\end{split}
\end{equation}
where $\sigma_{I=1,2,3}$ is the Pauli matrices. By \eqref{GammaMat}, the equation of motion \eqref{DiracEq3} can be further simplified as
\begin{equation} \label{}
\begin{split}
\sqrt{\frac{g_{xx}}{g_{rr}}}\left(\partial_{r} \mp m \sqrt{g_{rr}} \right)\phi_{\pm} \pm \left(ik \sigma_{1} + u \sigma_{2} \pm p \sigma_{2}\sqrt{\frac{g_{xx}}{g_{tt} g_{rr}}}A_{t}' \right)\phi_{\mp} = 0 \,.
\end{split}
\end{equation}
In addition, by introducing $\phi _{\pm}:=\left(\begin{array}{c}y_{\pm} \\z_{\pm}\end{array}\right)$, the above equation of motion becomes two decoupled equations:
\begin{equation} \label{DecouEq}
\begin{split}
\sqrt{\frac{g_{xx}}{g_{rr}}}\xi_{\pm}'   +  2m\sqrt{g_{xx}}\xi_{\pm}  - \left( u-p\sqrt{\frac{g_{xx}}{g_{tt}g_{rr}}}A_{t}' \mp k\right)   -  \left(u +p\sqrt{\frac{g_{xx}}{g_{tt} g_{rr}}}A_{t}' \pm k \right)\xi_{\pm}^2   = 0 \,,
\end{split}
\end{equation}
where 
\begin{equation} \label{xidef}
\xi_{+}:= i \frac{y_{-}}{z_{+}}\,,  \qquad \xi_{-}:= -i \frac{z_{-}}{y_{+}} \,.
\end{equation}
This decoupled equations \eqref{DecouEq} is called the flow equation and it can be solved analytically near the boundary ($r \rightarrow \infty$):
\begin{equation} \label{DecouEq2}
\begin{split}
\left(\begin{array}{c}y_{-} \\z_{+}\end{array}\right) = \left(\begin{array}{c}D_{1}\,r^{-m}+\tilde{A}_{1}\,r^{m-1} \\A_{1}\,r^m+\tilde{D}_{1}\,r^{-m-1}\end{array}\right)        \,, \quad
\left(\begin{array}{c}z_{-} \\y_{+}\end{array}\right) = \left(\begin{array}{c}D_{2}\,r^{-m}+\tilde{A}_{2}\,r^{m-1} \\A_{2}\,r^m+\tilde{D}_{2}\,r^{-m-1}\end{array}\right)  \,,
\end{split}
\end{equation}
with the  coefficients
\begin{equation} \label{DecouEq3}
\begin{split}
\tilde{A}_{1} &= \frac{i(k-\omega -q \mu)}{2m-1} A_{1} \,, \quad \tilde{D}_{1} = \frac{i(k+\omega +q \mu)}{2m+1} D_{1}\,, \\
\tilde{A}_{2} &= \frac{i(k+\omega +q \mu)}{2m-1} A_{2} \,, \quad \tilde{D}_{2} = \frac{i(k-\omega -q \mu)}{2m+1} D_{2}\,, \\
\end{split}
\end{equation}
where $\mu$ is the leading term of the gauge field, interpreted as a chemical potential in holography.
Note that the coefficients $\tilde{A}_{I=1,2}$ and $\tilde{D}_{I=1,2}$ in \eqref{DecouEq2} are determined by $A_{I=1,2}$ and $D_{I=1,2}$ through \eqref{DecouEq3}.
In other words, $(A_{1}, D_{1})$ and $(A_{2}, D_{2})$ in \eqref{DecouEq2} can be the two set of independent coefficients. 

Now we will formulate the retarded Green's function with a set of coefficients ($A_{I=1,2}$, $D_{I=1,2}$).
By following the prescription given in~\cite{Edalati:2010ww, Li:2011sh, Faulkner:2009wj, Liu:2009dm, Gubser:2009dt, Seo:2018hrc, Plantz:2018tqf}, if the fermion mass is in the range, $-1/2 < m < 1/2$~\cite{Iqbal:2009fd, Laia:2011wf},  we can develop the Green's function in two ways. One is called the standard quantization, and the other is the alternative quantization.  In the standard quantization, $A_{I}$ and $D_{I}$ are considered as the source and the corresponding response, respectively. On the other hand in the alternative quantization, the role of $A_{I}$ and $D_{I}$ is changed. i.e., $D_{I}$ is the source term and $A_{I}$ is the response term.

In the standard quantization, the retarded Green's function is given by
\begin{equation} \label{Standard}
\begin{split}
G^{R}(\omega, k \,; m,p) &= \text{diag}\left( i \frac{D_{1}}{A_{1}}, -i \frac{D_{2}}{A_{2}} \right) := \text{diag}\left(G_{+}^{R}, G_{-}^{R}\right) \,.
\end{split}
\end{equation}
By using \eqref{xidef}, \eqref{DecouEq2} and \eqref{DecouEq3},
the Green's function \eqref{Standard} can be expressed as
\begin{equation} \label{SF}
\begin{split}
G_{\pm}^{R}(\omega, k \,;m,p) &=  \frac{2m+1}{2m-1} \frac{(\pm k-q\mu-\omega)r^{2m} + (2m-1)r^{2m+1} \, \xi_{\pm}(\omega, k \,; m,p)}{(2m+1)r - (\pm k+q\mu+\omega)\xi_{\pm}(\omega, k \,; m,p)}  \,,
\end{split}
\end{equation}
which reduce to $G_{\pm}^{R}(\omega, k \,;m,p)=r^{2m} \xi_{\pm}$ at the boundary. 
Now we can construct the  spectral function which is defined as the imaginary part of the retarded Green's function:
\begin{equation} \label{SpectralStandard}
\begin{split}
A(\omega, k \,; m,p) := \text{Im}[\text{Tr}\, G^{R}(\omega, k \,;m,p)] \,.
\end{split}
\end{equation}

In the alternative quantization, the retarded Green's function can be written as
\begin{equation} \label{Standard2}
\begin{split}
\tilde{G}^{R}(\omega, k \,; m,p) &= \text{diag}\left( i \frac{A_{1}}{D_{1}}, -i \frac{A_{2}}{D_{2}} \right) := \text{diag}\left(\tilde{G}_{+}^{R}, \tilde{G}_{-}^{R}\right) \,.
\end{split}
\end{equation}
Similarly to \eqref{SF}, we have
\begin{equation} \label{SF2}
\begin{split}
\tilde{G}_{\pm}^{R}(\omega, k \,;m,p) =  \frac{-2m+1}{-2m-1} \frac{(\mp k-q\mu-\omega)r^{-2m} + (-2m-1)r^{-2m+1} \, \tilde{\xi}_{\pm}(\omega, k \,; m,p)}{(-2m+1)r - (\mp k+q\mu+\omega)\tilde{\xi}_{\pm}(\omega, k \,; m,p)} \,, 
\end{split}
\end{equation}
where $\tilde{\xi}_{\pm}(\omega, k \,; m,p) :=  -1/\xi_{\pm}(\omega, k \,; m,p)$ {and they reduce to $\tilde{G}_{\pm}^{R}(\omega, k \,;m,p)=r^{-2m} \tilde{\xi}_{\pm}$ at the boundary.}

It can be shown that there are the following relations between the retarded Green's function in the standard quantization and in the alternative quantization:
\begin{align} \label{}
G_{\pm}^{R}(\omega, k \,; m,p) &= \frac{-1}{\tilde{G}_{\pm}^{R}(\omega, k \,; m,p)}  \,, \quad G_{\pm}^{R}(\omega, -k \,; m,p) = G_{\mp}^{R}(\omega, k \,; m,p) \,, \\
\tilde{G}_{\pm}^{R}(\omega, k \,; m,p) &= G_{\pm}^{R}(\omega, -k \,; -m,-p) = G_{\mp}^{R}(\omega, k \,; -m,-p) \,, \label{Relation}
\end{align}
where $ \tilde{\xi}_{\pm}(\omega, k \,; m,p) = \xi_{\pm}(\omega, -k \,; -m,-p) $ is used.

Therefore, the Green's function with the negative fermion mass  in the standard quantization can be seen as the Green's function with the positive fermion mass  in the alternative quantization. 
It has been shown that the holographic spectral function $A$ in \eqref{SpectralStandard} behaves as $A \sim \omega^{2m}$ in the high frequency limit in the standard quantization~\cite{Gursoy:2011gz}. It means that $A \sim \omega^{-2m}$ in the alternative quantization. To have decreasing spectral functions at high frequencies, we will consider the positive fermion mass in the alternative quantization framework: $\tilde{G}_{\pm}^{R}(\omega, k \,; m, p)$.

\section{Gubser-Rocha-linear axion model}\label{ADM123}
Let us first introduce the Gubser-Rocha-linear axion model.~\cite{Davison:2013txa, Kim:2017dgz,Zhou:2015qui} This model is an extended version of the Gubser-Rocha model~\cite{Gubser:2009qt} by adding `axion' fields. 
The action is
\begin{equation}\label{axionmodel}
S=\int \mathrm{d}^4x\sqrt{-g} \left[R-\frac{1}{4} e^\phi F^2 -\frac{3}{2}(\partial{\phi})^2+\frac{6}{L^2}\cosh \phi -\frac{1}{2}\sum_{I=1}^{2}(\partial \psi_{I})^2  \right] \,,
\end{equation}
which consists of four fields: metric $g_{\mu\nu}$, U(1) gauge field $A_{\mu}$ and two scalar fields: `dilaton' field $\phi$ and `axion' field $\psi_{I=1,2}$. $L$ is the AdS radius. 
We will call this model `Gubser-Rocha-linear axion model', which can be considered as a minimum holographic model for condensed matter systems in the following sense.
The finite temperature and chemical potential (or density) are holographically related to the Hawking temperature of a black hole and the temporal gauge field $A_t$ at the AdS boundary. 
The dilaton field with a specific potential ($\cosh(\phi)$) is introduced to make entropy vanish at zero temperature: entropy density is linearly proportional to the temperature~\cite{Gubser:2009qt, Davison:2013txa}.
The axion field breaks the translational symmetry so renders the conductivity finite~\cite{Andrade:2013gsa, Davison:2013txa, Zhou:2015qui, Kim:2017dgz, Jeong:2018tua}. 

%
The action \eqref{axionmodel} yields the equations of motion
\begin{equation}\label{EiensteinEq}
\begin{split}
&R_{\mu\nu} -\frac{1}{2}g_{\mu\nu}\left[R-\frac{1}{4} e^\phi F^2 -\frac{3}{2}(\partial{\phi})^2+\frac{6}{L^2}\cosh \phi -\frac{1}{2}\sum_{I=1}^{2}(\partial \psi_{I})^2     \right] \\
& \qquad =\frac{1}{2}e^\phi F_{\mu\d}F_{\nu}{^\d}+\frac{3}{2}\partial_{\mu}\phi \partial_{\nu}\phi+\frac{1} {2}\sum_{I=1}^{2}(\partial_{\mu}\psi_{I}\partial_{\nu}\psi_{I}) \,, \\
&\grad^2\phi-\frac{1}{12}e^\phi F^2+\frac{2}{L^2} \sinh (\phi) =0 \,,\qquad  \grad_{\mu}(e^\phi F^{\mu\nu})=0  \,, \qquad \grad^{2}\psi_{I}=0 \,,
\end{split}
\end{equation}
which are satisfied by the following solution:
\begin{equation} \label{ansatz1}
\begin{split}
&\dd s^2 =  -g_{tt} \dd t^2  + g_{rr} \dd r^2 + g_{xx} \dd x^2 + g_{yy}\dd y^2 \\
& \quad \ \, = -\frac{\tilde{r}^2 g(\tilde{r}) h(\tilde{r})}{L^{2}} \, \mathrm{d} \tilde{t}^2 + \frac{L^2}{\tilde{r}^2 g(\tilde{r})h(\tilde{r})} \,\mathrm{d} \tilde{r}^2  + \frac{\tilde{r}^2 g(\tilde{r})}{L^2} \, \mathrm{d} \tilde{x}^2 + \frac{\tilde{r}^2 g(\tilde{r})}{L^2} \, \mathrm{d} \tilde{y}^2  \,,\\
&  \quad h(\tilde{r})=1-\frac{L^4 \tilde{\beta}^2}{2(1+\tilde{r})^2}-\frac{(1+\tilde{r}_h)^3}{(1+\tilde{r})^3}\left(1-\frac{L^4 \tilde{\beta}^2}{2(1+\tilde{r}_h)^2}\right) \,,  \quad g(\tilde{r})=\left(1+\frac{1}{\tilde{r}}\right)^\frac{3}{2} \,, \\
& A = \sqrt{\frac{3(1+\tilde{r}_h)}{L^2}\left(1-\frac{L^4 \tilde{\beta}^2}{2(1+\tilde{r}_h)^2}\right)}\left(1-\frac{1+\tilde{r}_h}{1+\tilde{r}}\right) \mathrm{d} \tilde{t} \,,  \\
&\phi=\frac{1}{3} \log(g(\tilde{r})) \,, \quad  \psi_1 =  \tilde{\beta} \, \tilde{x} \,, \quad  \psi_2 =  \tilde{\beta} \, \tilde{y} \,, 
\end{split}
\end{equation}
where 
\begin{equation} \label{tildes}
\tilde{r} :=  \frac{r}{Q} \,, \qquad \tilde{t} := t \, Q \,, \qquad \tilde{x} := x \, Q \,, \qquad \tilde{y} := y \, Q \,, \qquad  \tb :=  \frac{\beta}{Q} \,.
\end{equation}
Here, $\tilde{r}_{h}$ is the black hole horizon radius satisfying $h(\tilde{r}_{h})=0$ and $\tilde{\beta}$ is the strength of momentum relaxation or translational symmetry breaking.\footnote{{Note that the background solution \eqref{ansatz1} corresponds to the case with ~\cite{Kim:2017dgz}  by the transformation \eqref{tildes}. }}
For simplicity, we will set $L=1$ from here.

The temperature and chemical potential read, from \eqref{ansatz1},
\begin{align} 
    T &=  \frac{g_{tt}'}{4\pi\sqrt{g_{tt}g_{rr}}}\Bigr|_{r_{h}} = Q \, \frac{\sqrt{\tilde{r}_{h}}(6(1+\tilde{r}_{h})^2 - \tilde{\beta}^2)}{8\pi(1+\tilde{r}_{h})^{3/2}} =: Q \, \tilde{T}   \,, \label{T2} \\
\mu &=  A_{t}(r\rightarrow\infty)  = Q \sqrt{3(1+\tilde{r}_h)\left(1-\frac{\tilde{\beta}^2}{2(1+\tilde{r}_h)^2}\right)}  =: Q \, \tilde{\mu}  \label{mu2} \,,
\end{align}
where $\tilde{T} := T/Q$ and $\tilde{\mu} := \mu/Q$ are used in the last equalities.
In this paper, we will explore the spectral function at fixed chemical potential. 
For this purpose we define 
\begin{align}
&\bT := \frac{T}{\mu}      = \frac{\tT}{\tmu} =   \frac{ \sqrt{\tilde{r}_{h}} (6(1+\tilde{r}_{h})^2  - \tilde{\beta}^2)  }{ 4\pi(1+\tilde{r}_{h})\sqrt{12(1+\tilde{r}_{h})^2-6\tilde{\beta}^2} }            \,,  \label{bteq} \\ 
&\bb := \frac{\beta}{\mu} = \frac{\tb}{\tmu} =  \sqrt{\frac{2(1+\tilde{r}_{h})\tilde{\beta}^{2}}{3(2(1+\tilde{r}_{h})^2-\tilde{\beta}^2)}} \,. \label{bbeq}
\end{align}
By using these relations, the background solution \eqref{ansatz1}, expressed in terms of ($\tilde{r}_{h}$, $\tilde{\beta}$), can be determined by  ($\bar{T}$, $\bar{\beta}$).
%
%
%

The spectral function \eqref{SpectralStandard} with \eqref{SF2} in the background \eqref{ansatz1} can be expressed as
\begin{align}\label{}
\begin{split}
A(\tilde{\omega}, \tilde{k} \,; m,p)   &:= \text{Im}[\text{Tr}\, \tilde{G}^{R}(\tilde{\omega}, \tilde{k} \,;m,p)] \,, \\ 
\tilde{G}_{\pm}^{R}(\tilde{\omega}, \tilde{k} \,;m,p) &=  \frac{-2m+1}{-2m-1} \frac{(\mp \tilde{k}-q\tilde{\mu}-\tilde{\omega})\tilde{r}^{-2m} + (-2m-1)\tilde{r}^{-2m+1} \, \tilde{\xi}_{\pm}(\tilde{\omega}, \tilde{k} \,; m,p)}{(-2m+1)\tilde{r} - (\mp \tilde{k}+q\tilde{\mu}+\tilde{\omega})\tilde{\xi}_{\pm}(\tilde{\omega}, \tilde{k} \,; m,p)} \,,
\end{split}
\end{align}
where $\tilde{\omega}:=\omega/Q$ and $\tilde{k}:=k/Q$ are used. Thus, the spectral function at fixed chemical potential reads
\begin{align}\label{}
\bar{A}(\bar{\omega}, \bar{k} \,; m,p)   &:= \frac{A(\tilde{\omega}, \tilde{k} \,; m,p)}{\tilde{\mu}^{-2m}}  \,,
\end{align}
where $\bar{\omega}:= \omega/\mu = \tilde{\omega}/\tilde{\mu}$ and $\bar{k}:= k/\mu = \tilde{k}/\tilde{\mu}$.

In summary, we have five parameters: ($m, p, q$) from the probe fermion action \eqref{SpinorAction} and ($\bar{T}, \bar{\beta}$) from the background action \eqref{ansatz1}. 
In this paper we will set $q=1$ and $\bar{T}=1/10$ and investigate various phases inferred from the fermionic spectral function $\bar{A}(\bar{\omega}, \bar{k})$ in terms of ($m, p, \bar{\beta}$).
First, we study phases without momentum relaxation, i.e. $\bar{\beta} = 0$, in section \ref{sec31}.  Next, we will see how momentum relaxations affect those phases in section \ref{sec32}.

\subsection{Zero momentum relaxation} \label{sec31}

The phase structure by the spectral function in the space of $(m, p)$ was first investigated in~\cite{Seo:2018hrc}, where the RN-AdS model was considered. Here, we closely follow the method of ~\cite{Seo:2018hrc}, but in the Gubser-Rocha-linear axion model. This model is phenomenologically more appealing since it has zero entropy at zero temperature and exhibits linear-$T$ resistivity in low and high temperature.

\begin{figure}[] 
\centering
     \subfigure[Fermi liquid like (FL)]
     {\includegraphics[width=4.83cm]{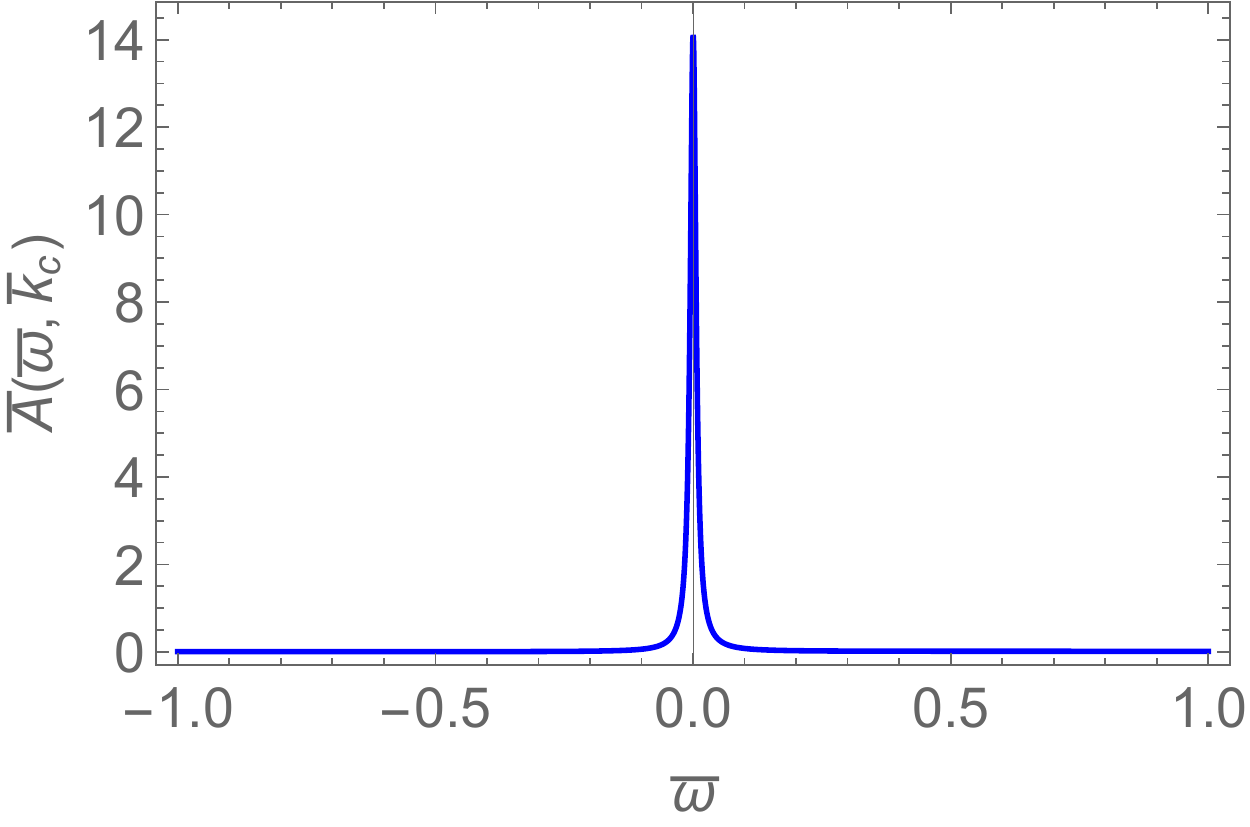} \label{FLLfig}}
           \subfigure[Bad metal prime (BM')]
     {\includegraphics[width=4.83cm]{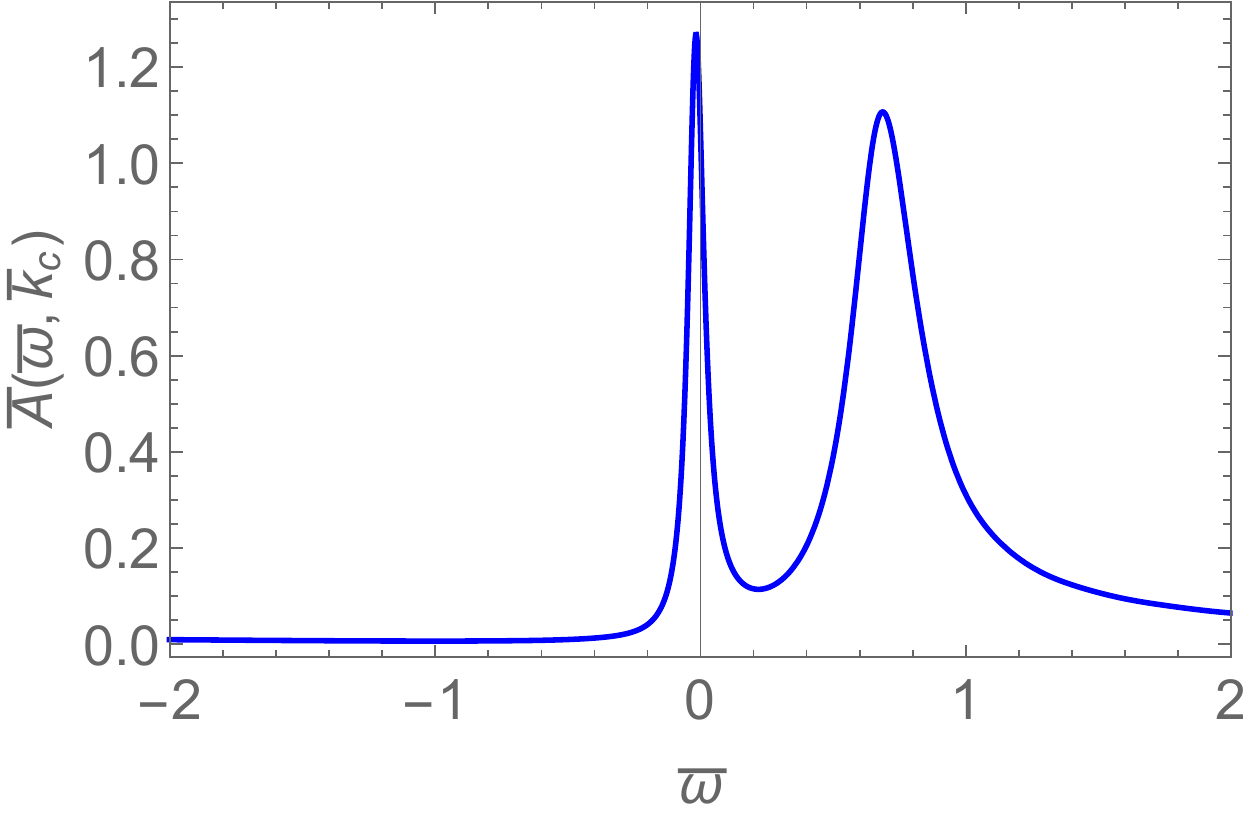} \label{Bmpfig}}
           \subfigure[Half-metal (hM)]
     {\includegraphics[width=4.83cm]{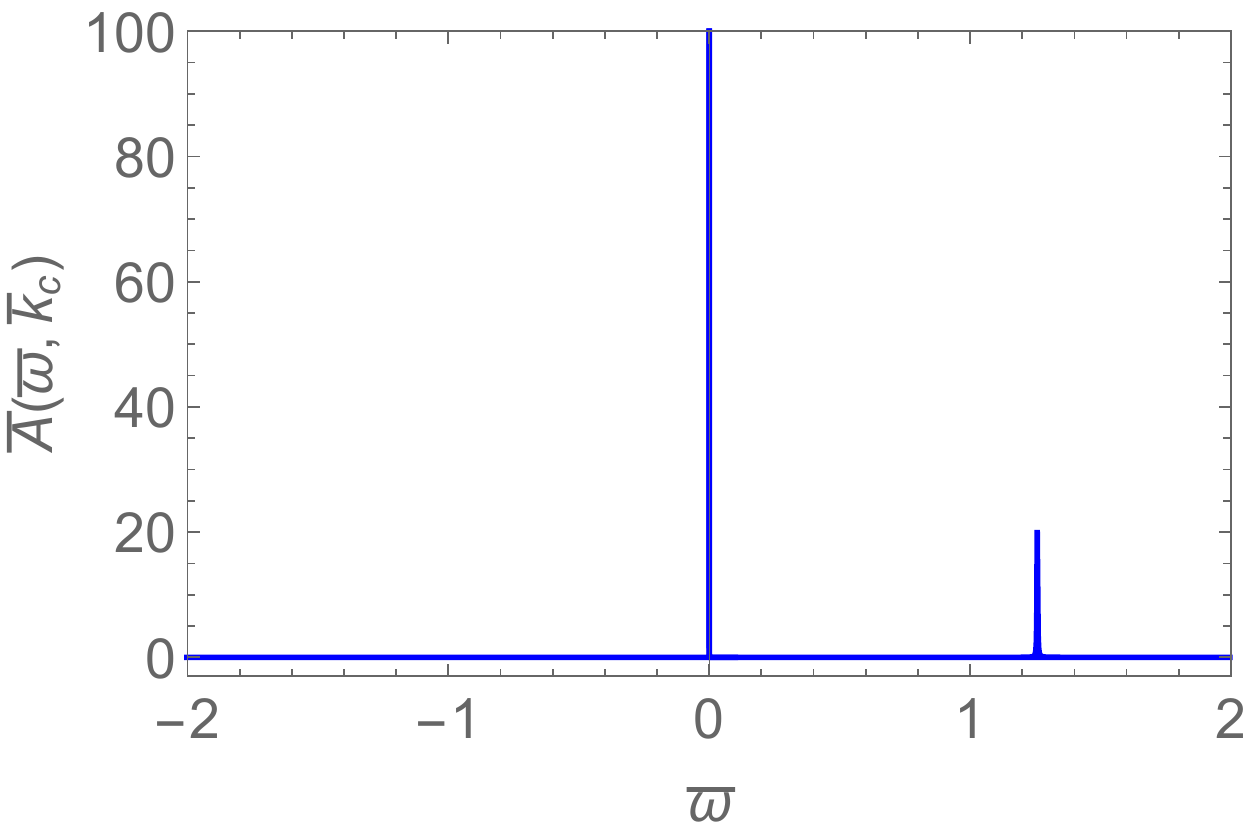} \label{hMfig}}
          \subfigure[Bad metal (BM)]
     {\includegraphics[width=4.83cm]{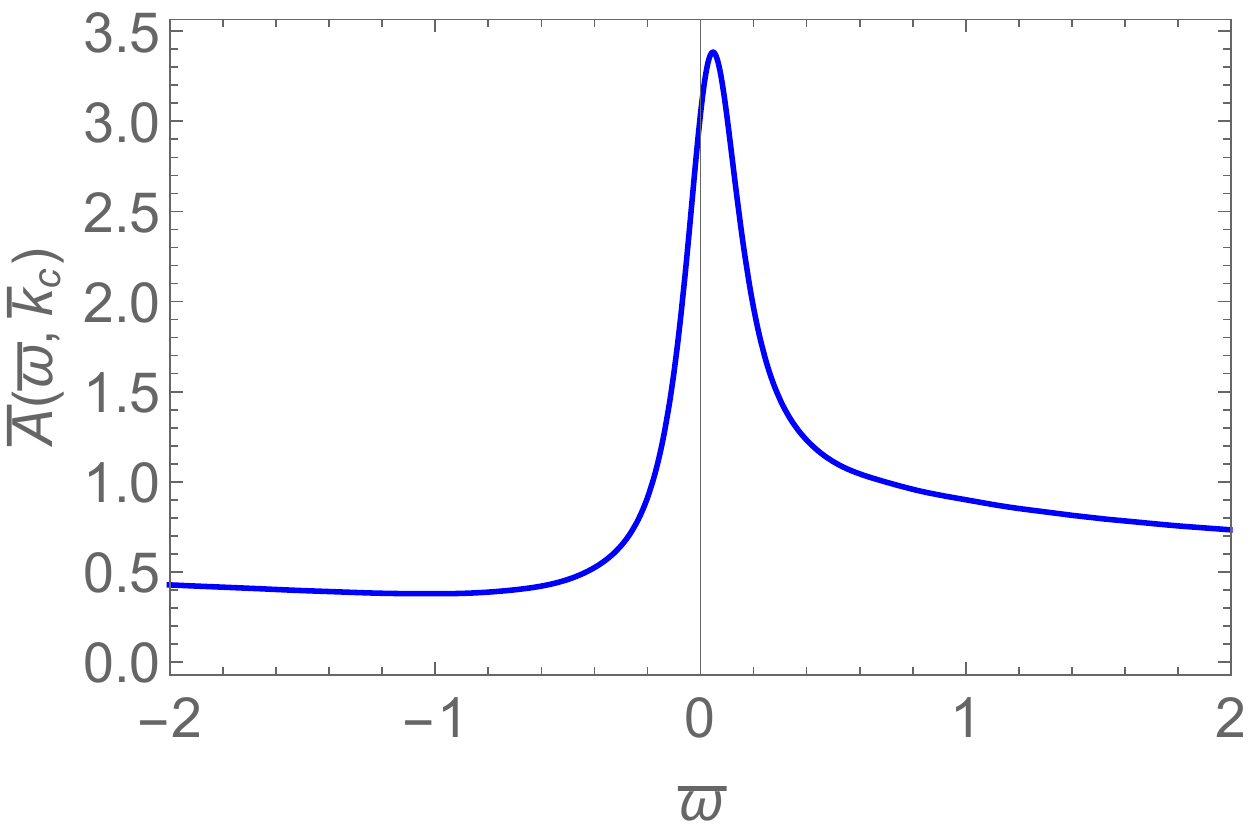} \label{BMfig}}
      \subfigure[Pseudogap (PG)]
     {\includegraphics[width=4.83cm]{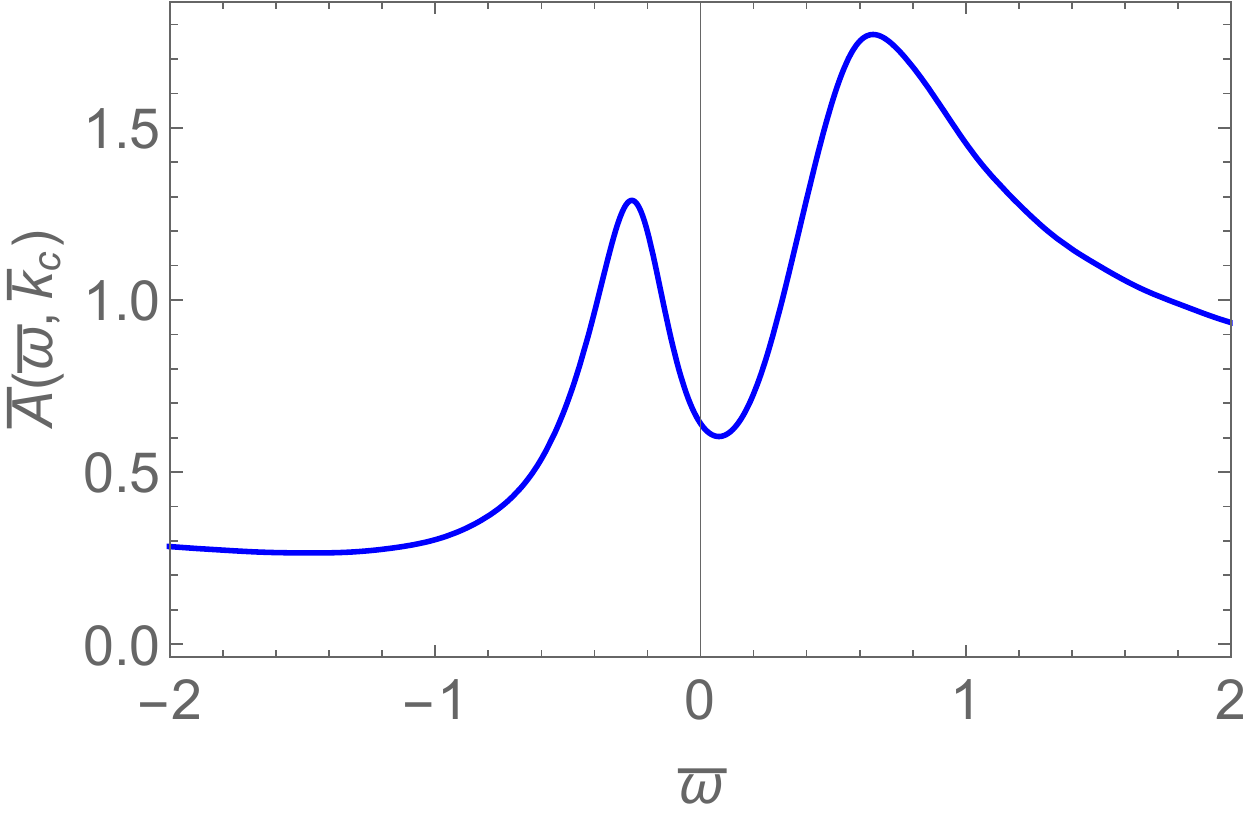} \label{PGfig}}
                \subfigure[Gapped (G)]
     {\includegraphics[width=4.83cm]{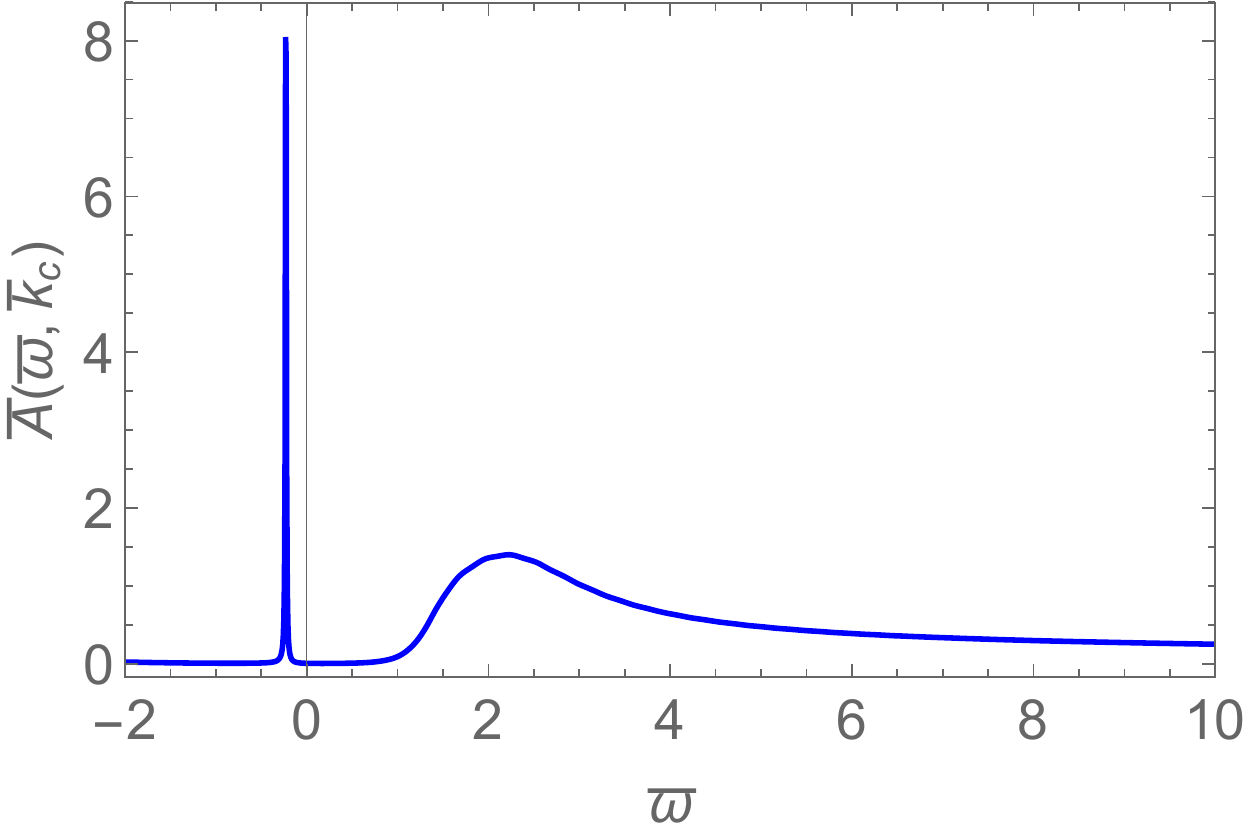} \label{Gfig}}
 \caption{Typical fermion phases without momentum relaxation. (a) FL with ($m$=0.45, $p=0.5$, $\bar{k}_{c}$=0.95),  (b) BM' with ($m$=0.40, $p=3.3$, $\bar{k}_{c}$=1.63), (c) hM with ($m$=0.45, $p=7$, $\bar{k}_{c}$=1.45), (d) BM with ($m$=0.10, $p=0.5$, $\bar{k}_{c}$=0.62),  (e) PG with ($m$=0.10, $p=2$, $\bar{k}_{c}$=1.2)  (f) G(Gapped phase) with ($m$=0.20, $p=7$, $\bar{k}_{c}$=3.3). The parameters ($m,p$) for these plots are shown as blue dots in Fig.~\ref{ZeroGravitonMass2}.  }\label{Typical}
\end{figure}

\begin{figure}[]
\centering
     \subfigure[$m=0.20$, $\bar{k}_{c}=3.3$:  gapped phase. ] 
     {\includegraphics[width=4.83cm]{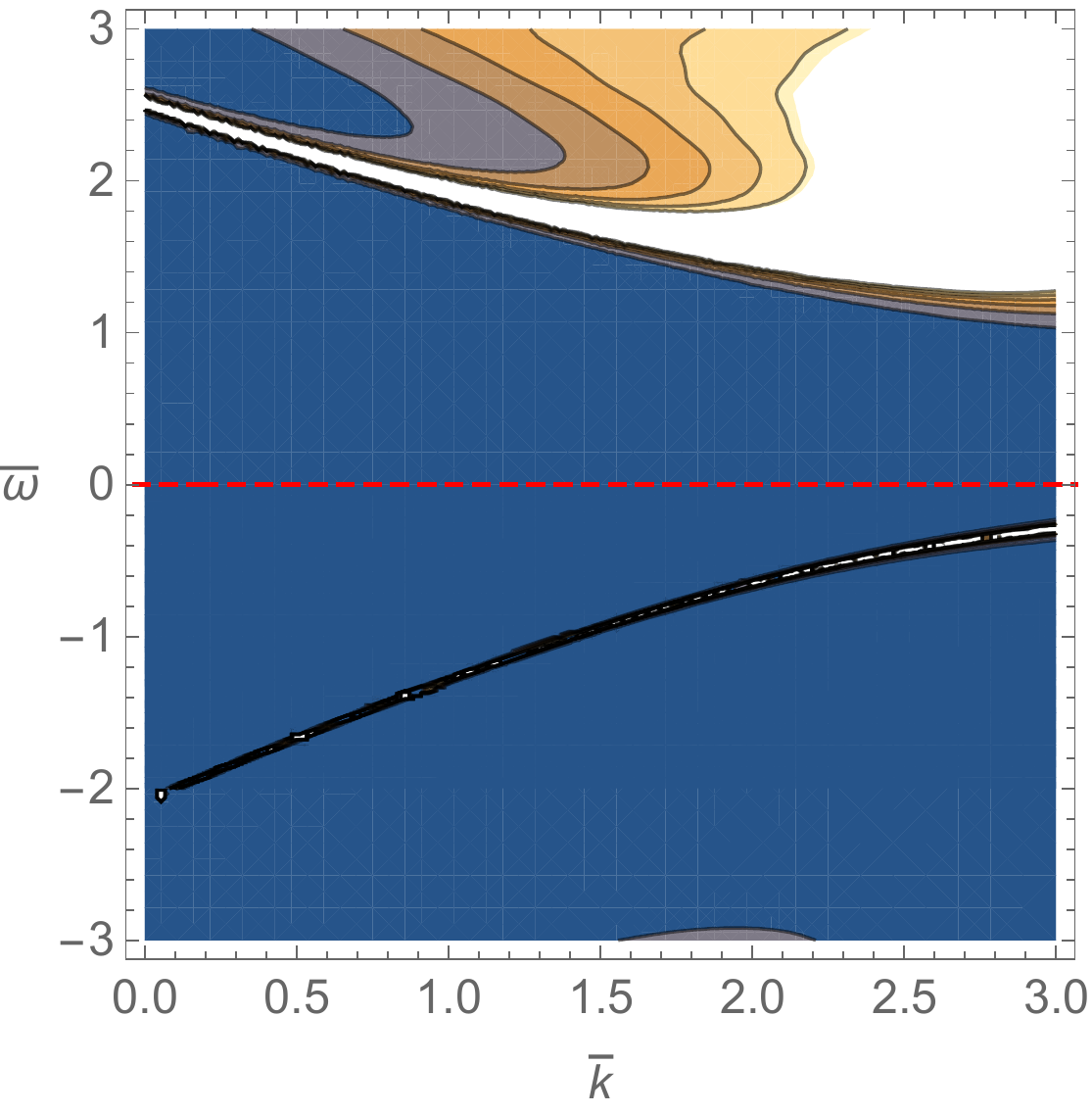} \label{hMMecha}}
     \subfigure[$m=0.35$, $\bar{k}_{c} = 3.3$]
     {\includegraphics[width=4.83cm]{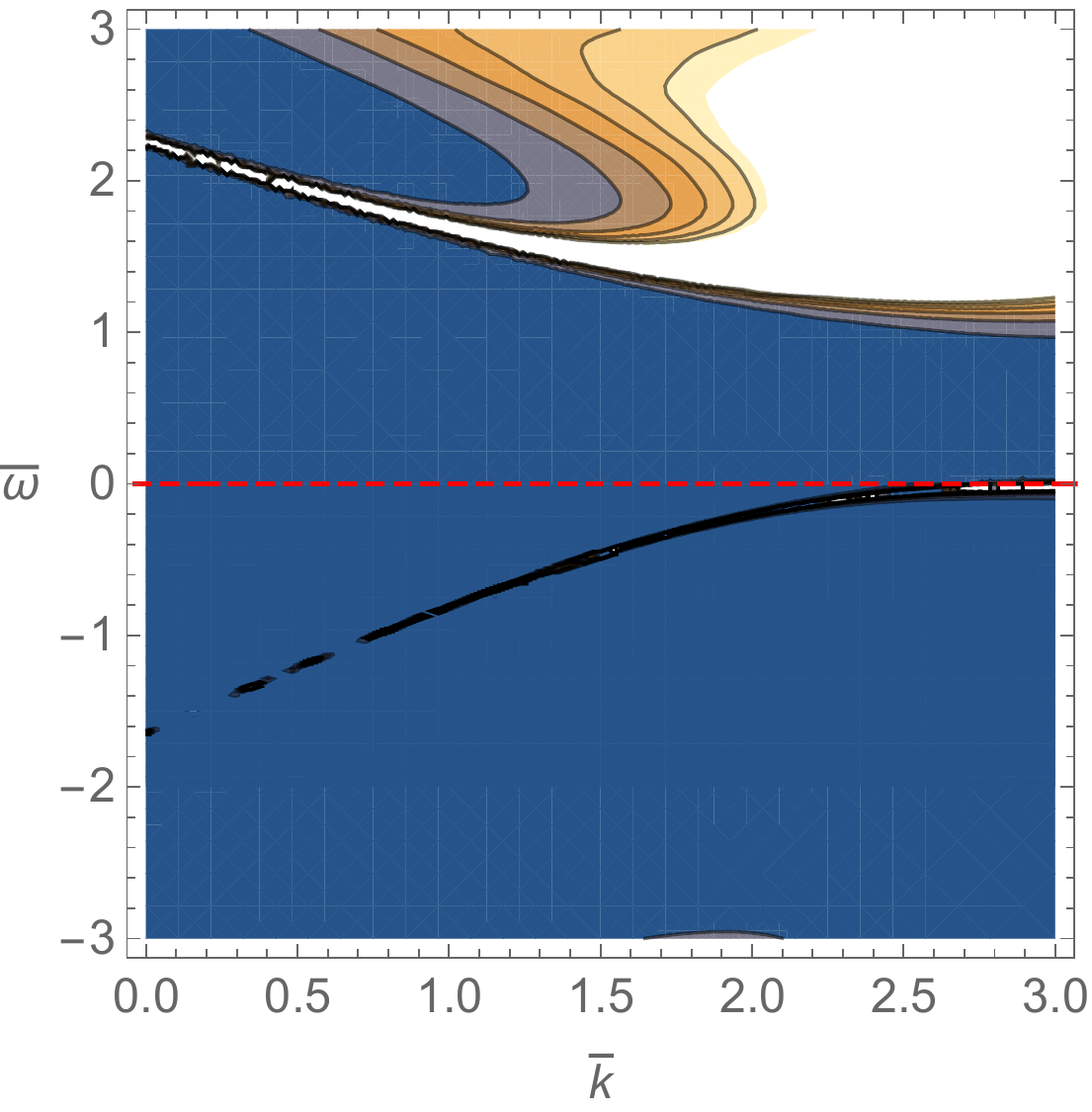} \label{hMMechb}}
     \subfigure[$m=0.45$, $\bar{k}_{c} =1.45$: half-metal phase.]
     {\includegraphics[width=4.83cm]{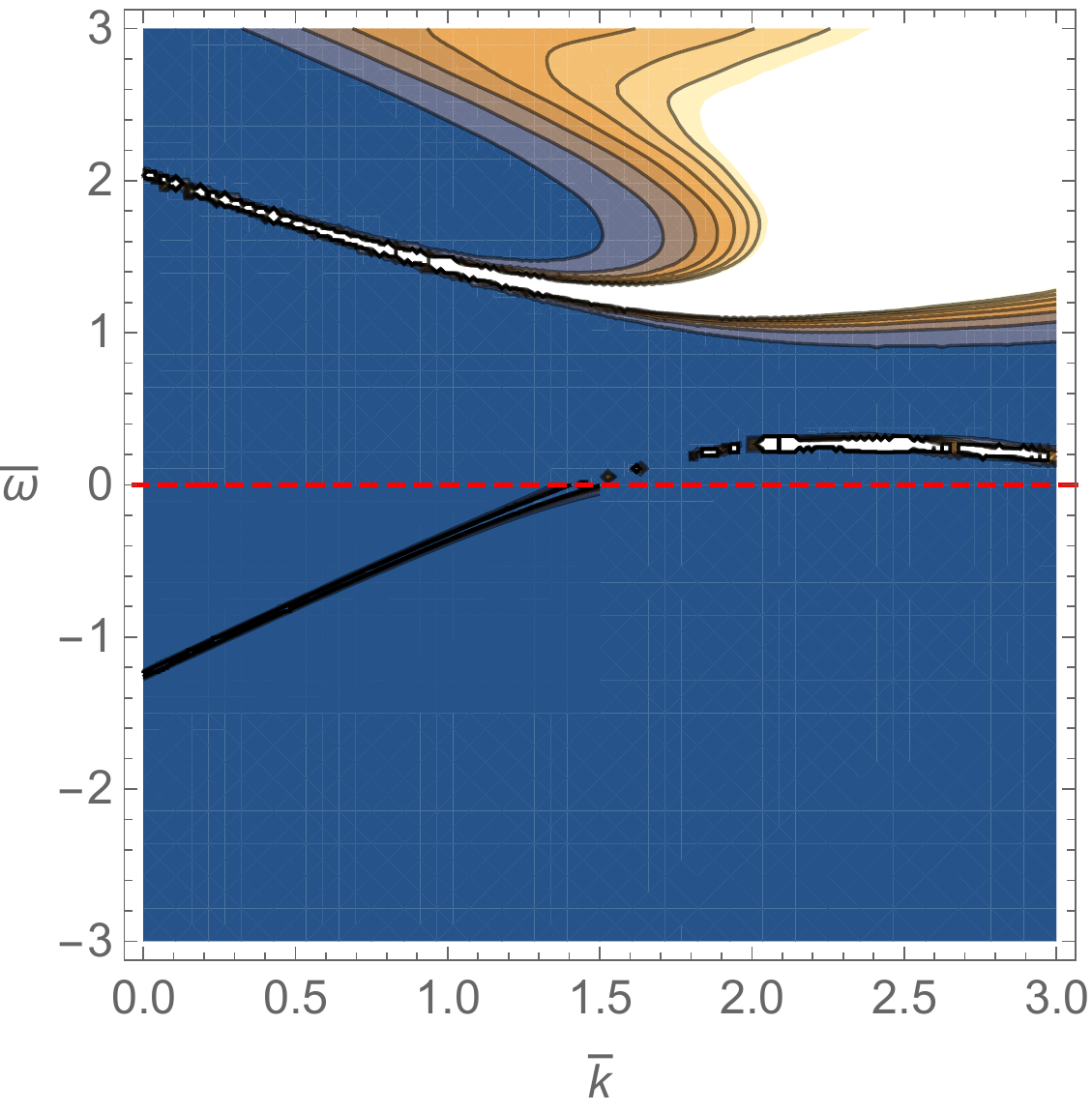} \label{hMMechc}}
 \caption{The density plot of the spectral function at $p=7$.}\label{hMMech}
\end{figure}

We display the typical  spectral function of our model, the Gubser-Rocha-linear axion model, in Fig.~\ref{Typical} to illustrate various phases of the holographic fermion. 
The spectral function in Fig.~\ref{Typical} can be understood as a cross section of Fig.~\ref{hMMech} at a fixed momentum $\bar{k}_{c}$. The $\bar{k}_c$ is the fermi momentum ($\bar{k}_{F}$) if there is a clear fermi surface. For example, see Fig. \ref{hMMechb} and \ref{hMMechc}, where $\bar{k}_F = 3.3,\, 1.45$ respectively. We will call this phase `gapless phase'. 
If there is no clear fermi surface, we choose to use $\bar{k}_c$ as the value when the dispersion curve is separated from $\bar{\omega}=0$ line. For example, see Fig. \ref{hMMecha}, where $\bar{k}_F=3.3$, which is taken from Fig. \ref{hMMechb}. Here, from Fig.~\ref{hMMechb} to Fig.~\ref{hMMecha} we changes $m$ with a fixed $p$. We call this phase `pseudogap phase' or `gapped phase': if the spectral function is non-zero at $\bar{\omega} =0$, it is a pseudogap phase (Fig.~\ref{PGfig}). Otherwise, it is a gapped phase (Fig.~\ref{Gfig}). Note that there are two peaks at finite $\bar{\omega}$ in these phases. 

The gapless phase is further divided as four sub-classes: fermi liquid like (FL), bad metal (BM), bad metal prime (BM'), half-metal (hM). 
FL (Fig. \ref{FLLfig}) has a single sharp and tall peak at the zero frequency. 
Because the spectral function has a finite peak, not a delta function peak we call it fermi liquid like. 
%
BM (Fig. \ref{BMfig}) also has a peak at zero $\bar{\omega}$ in the spectral function. However it has a broader peak compared with FL and goes to zero very slowly as $\bar{\omega}$ increases. 
BM' and hM have two peaks: one at $\bar{\omega}=0$ and the other at finite $\bar{\omega}$. If the spectral function is non-zero between two peaks, we call it BM'(Fig.~\ref{Bmpfig}.) while if it is zero, we call it hM (Fig. \ref{hMfig}).

\begin{figure}[]
\centering
     \subfigure[Phase diagram]
     {\includegraphics[width=7.3cm]{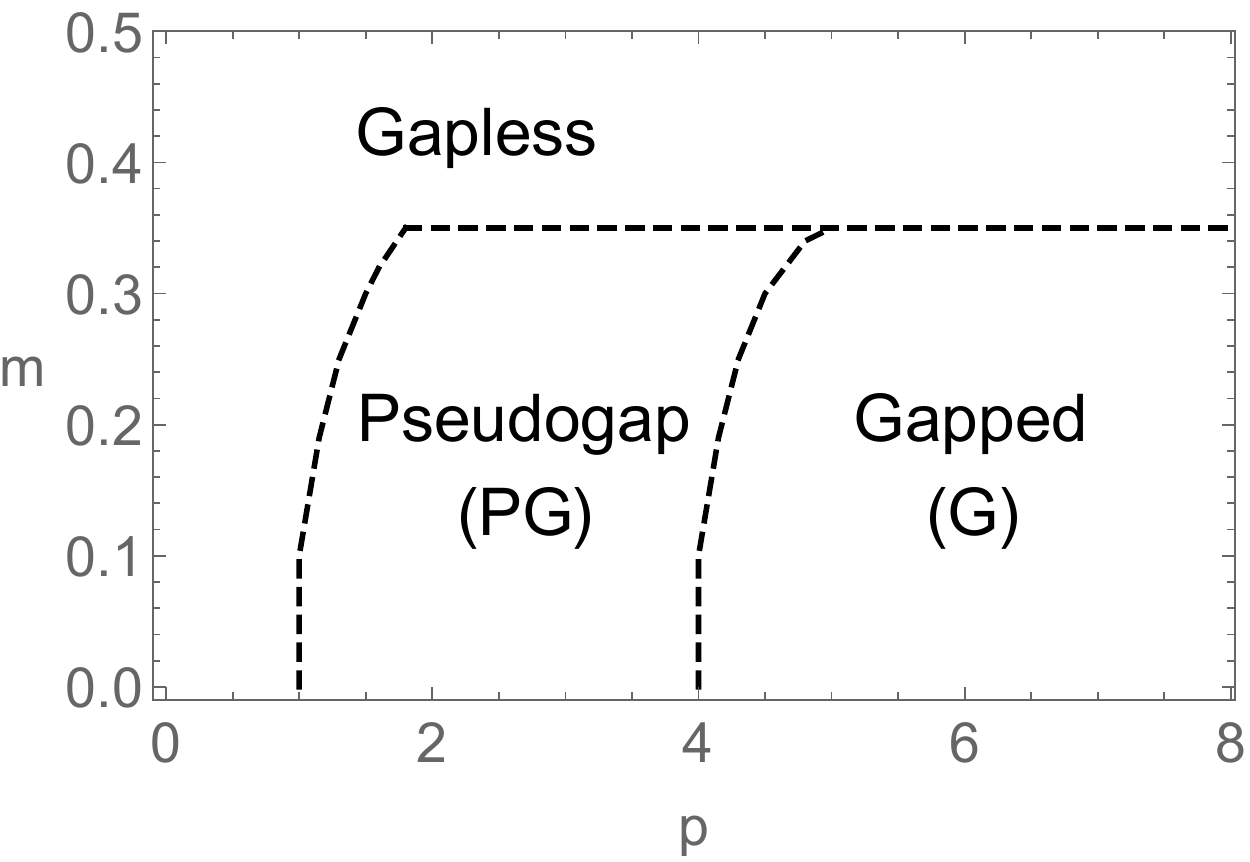} \label{ZeroGravitonMass1}}
     \subfigure[Substructure in gapless phase]
     {\includegraphics[width=7.3cm]{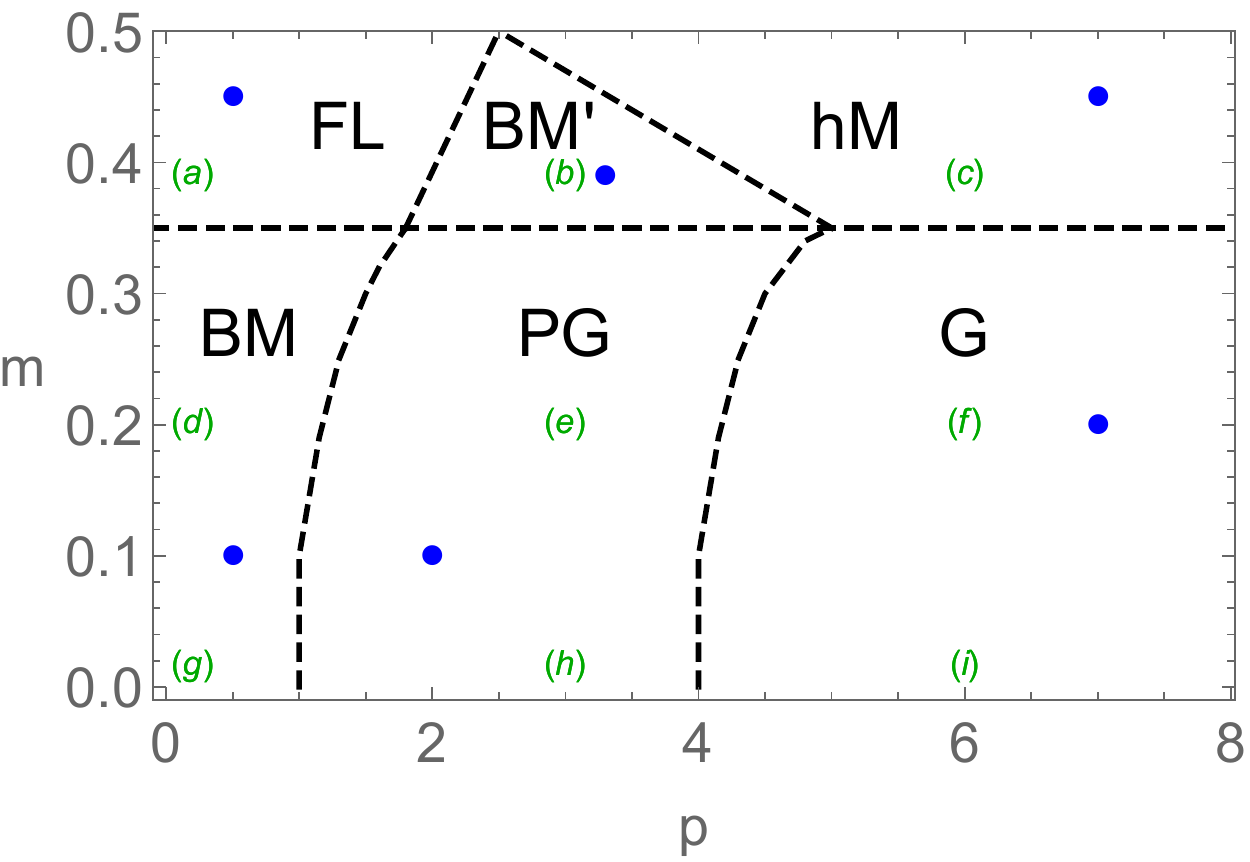} \label{ZeroGravitonMass2}}
 \caption{Phase diagram in ($m, p$) space without momentum relaxation.}\label{PhaseDiagram1}
\end{figure}

To make a phase diagram in the $(m, p)$ space, i) we choose one pair of $(m,p)$; ii) draw the density plot such as Fig.~\ref{hMMech}; iii) make a spectral function plot at $\bar{k}=\bar{k}_c$ such as Fig.~\ref{Typical}; iv) identify its phase according to the classification in Fig.~\ref{Typical}. Our final result is shown in Fig.~\ref{PhaseDiagram1}. For example, the blue points in Fig.~\ref{PhaseDiagram1} correspond to the plots in Fig.~\ref{Typical}. 
For the classification of the phases in Fig.~\ref{Typical} and Fig.~\ref{PhaseDiagram1}, we followed the conventions in~\cite{Seo:2018hrc}, where the same analysis has been done for the AdS-RN black hole case. Our result, Fig.~\ref{PhaseDiagram1}, is qualitatively the same as the RN-AdS case in~\cite{Seo:2018hrc}. 

For a fixed $m$, the dipole coupling has two effects. For $m>0.35$ it generates a new metalic phases: BM' and hM, while $m<0.35$ it generates a gap. For a larger $m$, a larger $p$ is required to open a gap. This gap generation is smooth via a pseudogap phase.  This supports the claim in~ \cite{Seo:2018hrc}: the existence of the smooth transition region between gapless phase and gapped phase is a general feature of strongly coupled system. Coming back to Fig.~\ref{Typical} with the understanding of Fig.~\ref{PhaseDiagram1} we note that: as $p$ increases, the spectral function develops another peak while the original peak first becomes broad but again becomes sharper. For a fixed $p$, as $m$ increases, the spectral function tends to be sharper and align its center to $\bar{\omega}=0$, which means gapless.

\subsection{Finite momentum relaxation} \label{sec32}

\begin{figure}[]
\centering
     {\includegraphics[width=15.3cm]{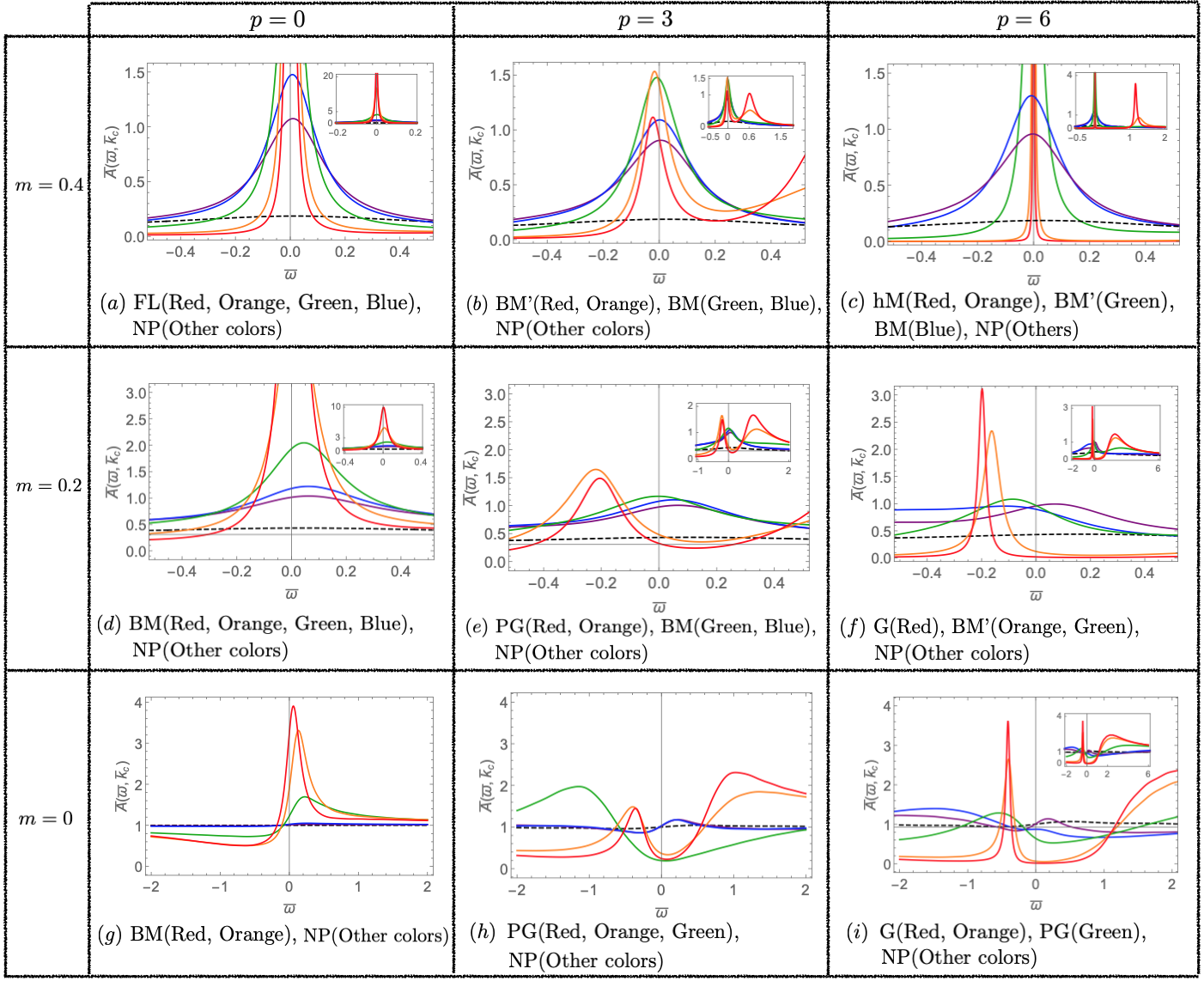} \label{}}
\caption{Spectral functions at finite momentum relaxation. Different color refers to different momentum relaxation: $\bar{\beta}$ = 0, 1, 3, 7, 10, 100 (red, orange, green, blue, purple, dashed black). }\label{m0FIG}
\end{figure}

Next, we investigate the effect of momentum relaxation on the spectral function. For illustrative purpose, we choose nine parameter points out of $m=(0,0.2,0.4)$ and $p=(0,3,6)$, which are marked as green letters (a),(b),$\cdots$(i) in Fig.~\ref{ZeroGravitonMass2}.
As momentum relaxation becomes stronger the spectral functions are changed as shown in Fig.~\ref{m0FIG}. 
\begin{figure}[]
\centering
     {\includegraphics[width=13cm]{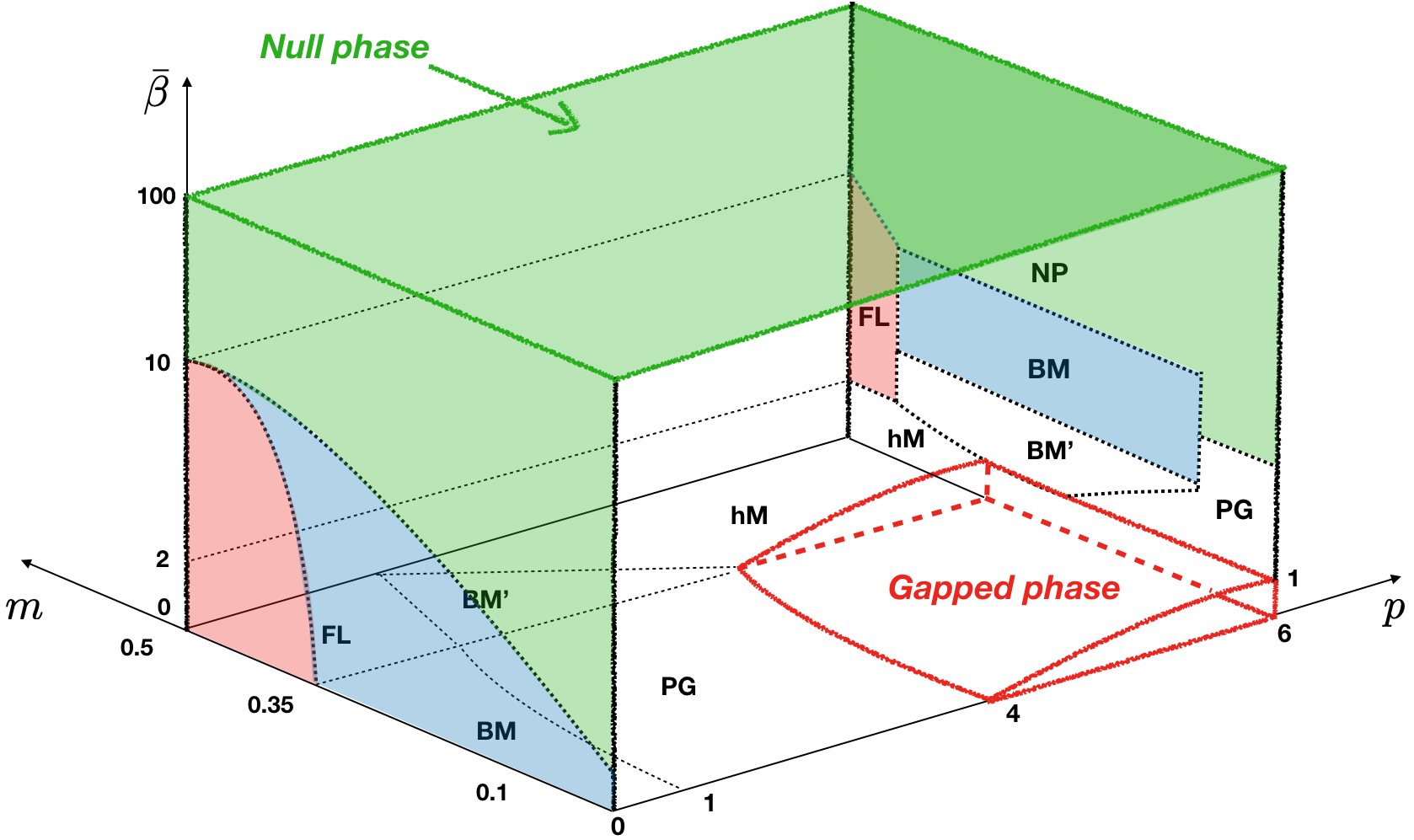} \label{}}
 \caption{Phase diagram in ($m, p, \bar{\beta}$) space. }\label{CompletePhaseD22}
\end{figure}
\begin{figure}[]
\centering
     \subfigure[$p=0$]
     {\includegraphics[width=4.82cm]{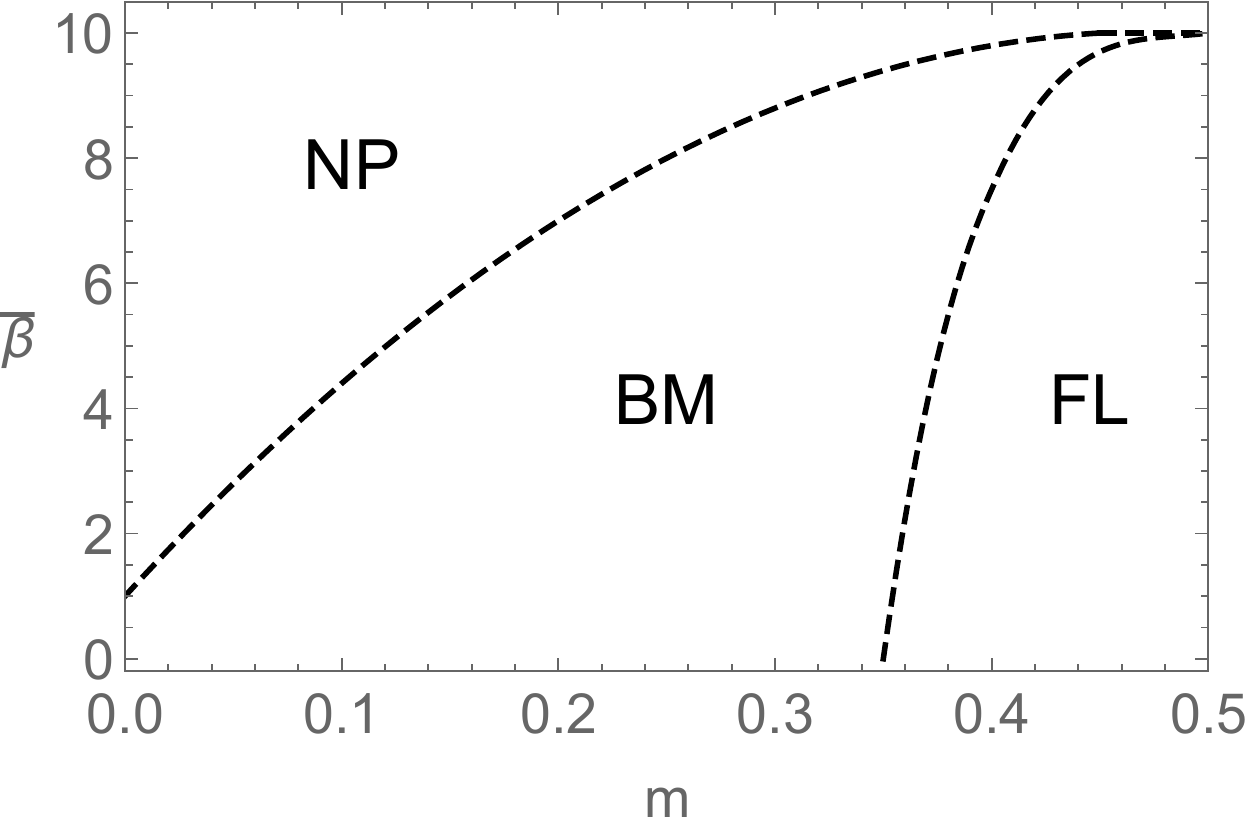} \label{FIG90}}
     \subfigure[$p=3$]
     {\includegraphics[width=4.82cm]{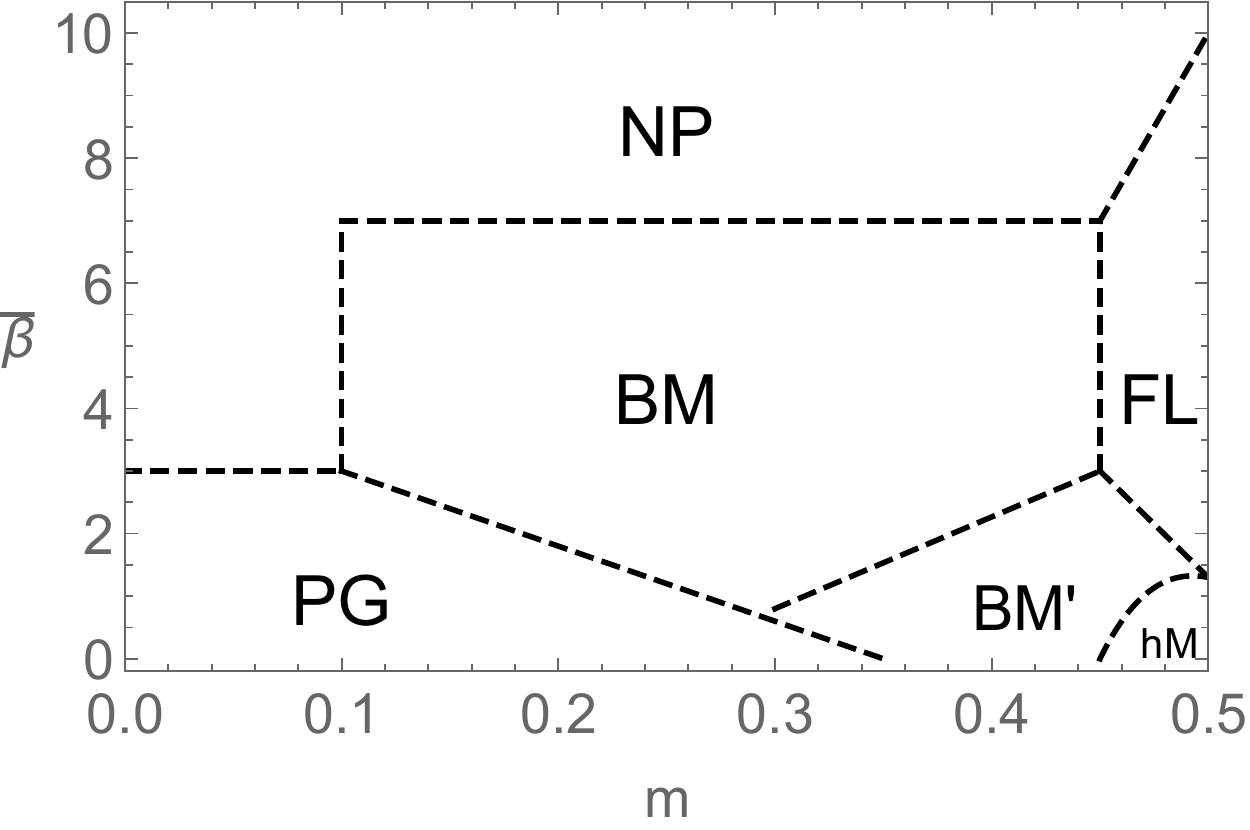} \label{FIG93}}
     \subfigure[$p=6$]
     {\includegraphics[width=4.82cm]{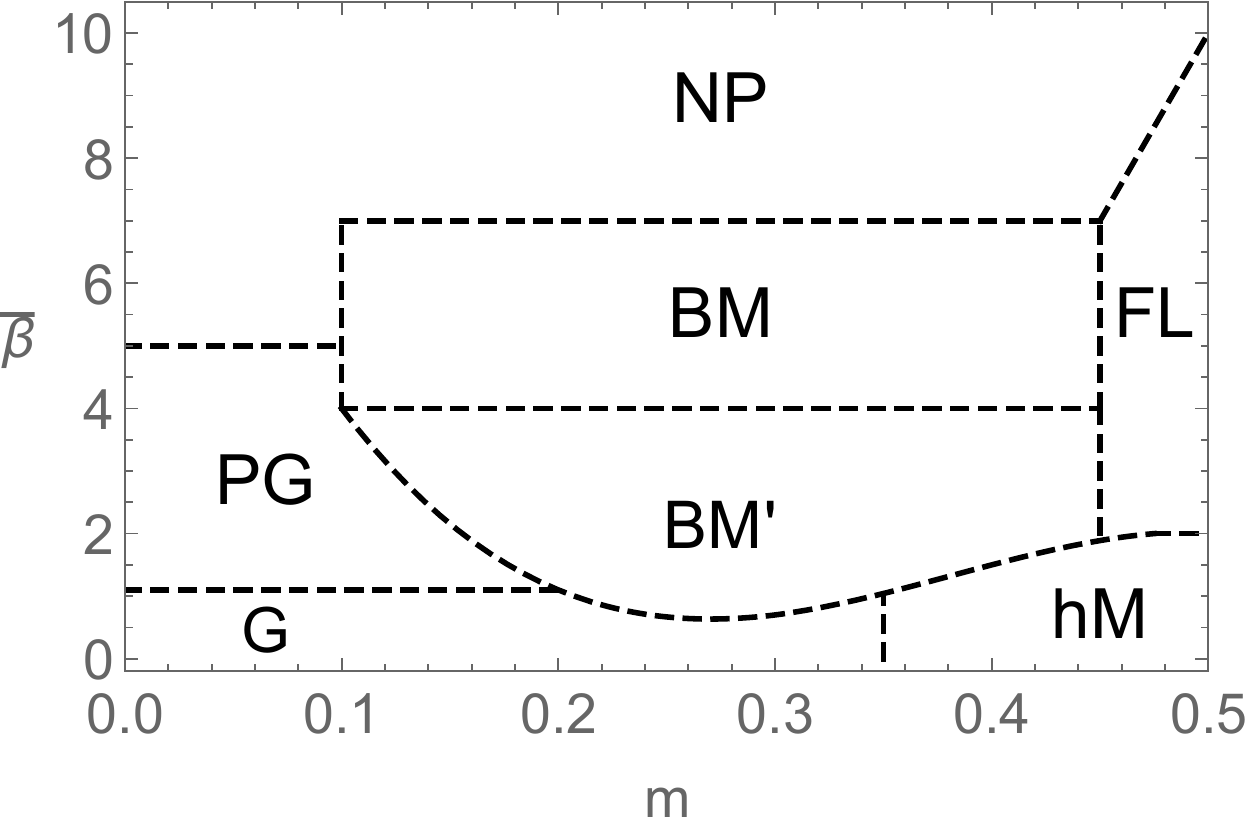} \label{FIG96}}
 \caption{Phase diagram in ($m$, $\bar{\beta}$) space. }\label{FIG9}
\end{figure}

The top row is for $m=0.4$, the middle row is for $m=0.2$, and the bottom one is for $m=0$. The left column is for $p=0$, the middle column is for $p=3$, and the right one is for $p=6$. The colors of the curves represent the strength of momentum relaxation, i.e. $\bar{\beta}$ = 0, 1, 3, 7, 10, 100 (red, orange, green, blue, purple, dashed black). 

Here, a new phase, null-phase (NP) is introduced for spectral functions at large momentum relaxation. This is an almost featureless phase. To quantify NP we have defined 
\begin{equation} \label{etaeq}
\begin{split}
\eta := \bar{A}_{\text{Highest}}(\bar{\omega}) - \bar{A}(\abs{\bar{\omega}} \gg 1),
\end{split}
\end{equation}
where $\bar{A}_{\text{Highest}}(\bar{\omega})$ is the highest value of the spectral function\footnote{For the numerical calculation, we evaluate $\bar{A}(\abs{\bar{\omega}} \gg 1)$ at $\abs{\bar{\omega}} = 10$.}. 
The NP is identified as a phase with $\eta<1$.

As momentum relaxation becomes strong the phase at every point in Fig.~\ref{m0FIG} changes as follows:
\begin{align}\label{CASES}
\begin{split}
&\text{(a): FL} \rightarrow \text{NP} \,, \qquad\qquad\quad\,\,\,  \text{(b): BM'} \rightarrow \text{BM} \rightarrow \text{NP} \,, \qquad\,\,\, \text{(c): hM} \rightarrow  \text{BM'} \rightarrow \text{BM} \rightarrow \text{NP} \,, \\
&\text{(d): BM} \rightarrow \text{NP} \,, \qquad\qquad\quad\,  \text{(e): PG} \rightarrow \text{BM} \rightarrow \text{NP} \,, \qquad\,\,\,\,\,\, \text{(f): G} \rightarrow  \text{BM'} \rightarrow  \text{NP} \,, \\
&\text{(g): BM} \rightarrow  \text{NP}  \,, \qquad\qquad\quad\,  \text{(h): PG} \rightarrow \text{NP} \,, \qquad\qquad\quad\,\,\,\,\,\,\, \text{(i): G} \rightarrow \text{PG} \rightarrow \text{NP} \,. \\
\end{split}
\end{align}
In this way, we can identify all phases in three dimensional space $(m,p,\bar{\beta})$. For example, see Fig.~\ref{CompletePhaseD22}, where to make our presentation clear we focus on some boundary cross-sections such as $p=0, 6$ and $\bar{\beta}=100$. For other cross-sections such as $p=3$ see Fig.~\ref{FIG9}.\footnote{If we put Fig. \ref{FIG93} to Fig. \ref{CompletePhaseD22}, it will be too complicated.}.

Note that, in Fig.~\ref{CompletePhaseD22} , the bounded red region corresponds to the gapped phase. The phase NP including strong momentum relaxation limit ($\bar{\beta}=100$) is plotted in green color. 
As momentum relaxation becomes stronger, 
the spectral function is suppressed in the whole range of ($m, p$)
and the various phases become featureless (NP) at strong momentum relaxation.

\begin{figure}[]
\centering
     {\includegraphics[width=7.5cm]{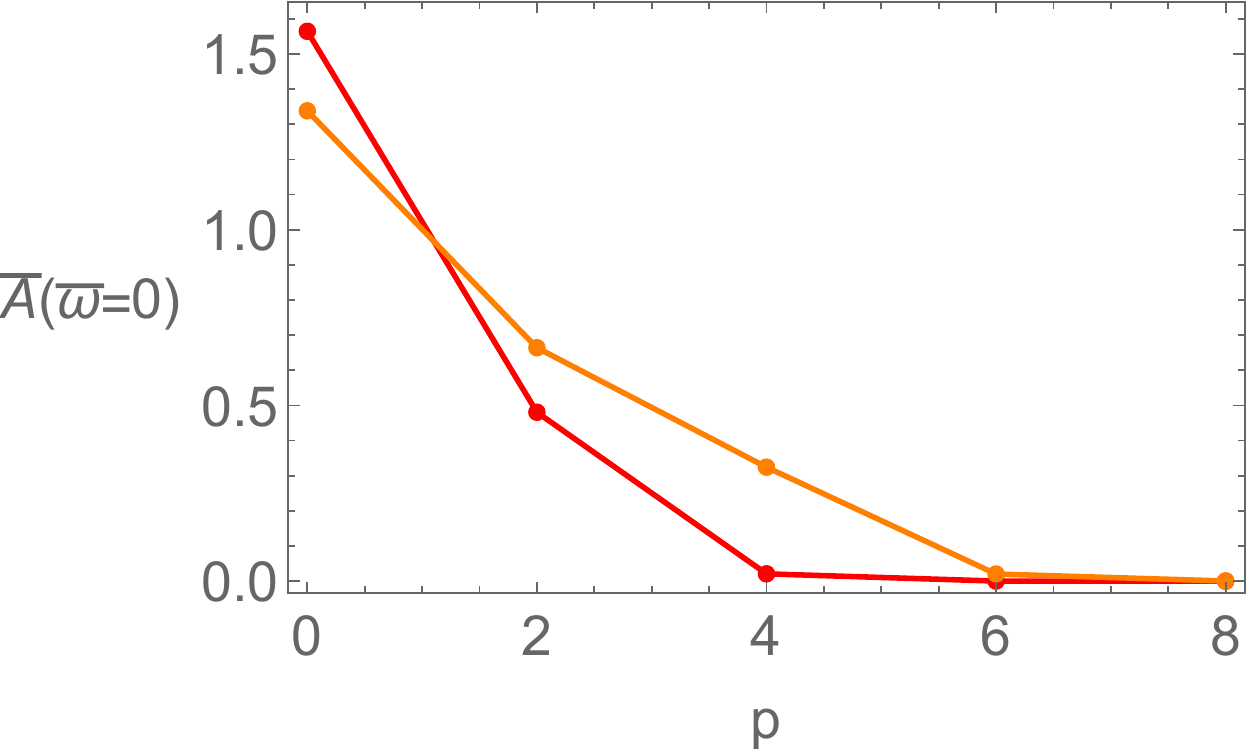} \label{}}
 \caption{Density of state at $\bar{\omega}=0$: the red is for $\bar{\beta}= 0$ and the orange is for $\bar{\beta}= 1$; ($m=0$).}\label{DOSPLOTT}
\end{figure}

As a cross-check of our phase diagram for gapped phases in Fig~\ref{FIG9}, we have also computed the density of state $\bar{A}(\bar{\omega})$, which is defined as
\begin{equation}
\bar{A}(\bar{\omega}) := \int \dd \bar{k} \, \bar{A}(\bar{\omega}, \bar{k}) \,.
\end{equation}
The gapped phase is identified by the vanishing density of state at $\bar{\omega}=0$ i.e. $\bar{A}(0) = 0$~\cite{Edalati:2010ww, Fang:2016wqe}. 
For example, for $m=0$, we show $\bar{A}(0)$ as a function of $p$ for $\bar{\beta} = 0,1$ in Fig.~\ref{DOSPLOTT}.
$\bar{A}(0)$ becomes zero\footnote{For the numerical calculation, we considered $\bar{A}(\bar{\omega}=0) \sim 10^{-2}$ is zero.} from $p=4$ for $\bar{\beta}=0$ and $p=6$ for $\bar{\beta}=1$. These agree with the values in Fig.~\ref{CompletePhaseD22}. We have also confirmed this agreement for different values of $m$.

%

\section{Linear axion model and comparison}
In this section, we study a 4-dimensional Einstein-Maxwell-axion model so called linear axion model~\cite{Andrade:2013gsa}.
It can be seen as an extension of \cite{Seo:2018hrc} to finite momentum relaxation cases.
The action is 
\begin{eqnarray}
S&=&\int
\mathrm{d}^4x\sqrt{-g} \left[ R+\frac{6}{L^2}-F^2 -\frac{1}{2} \sum_{I=1}^{2} (\partial \psi_I)^2 \right] \,,  \label{eq:action}
\end{eqnarray}
which consists of the metric $g_{\mu\nu}$, U(1) gauge field $A_{\mu}$,  and axion fields $\psi_{I=1,2}$.
Classical solutions are
\begin{equation}\label{LAMsol}
\begin{split}
\mathrm{d} s^2 &= -  f(r) \mathrm{d} t^2 +  \frac{\mathrm{d} r^2}{f(r)}  + \frac{r^2}{L^2}(\mathrm{d} x^2 + \mathrm{d} y^2)\,, \\ 
& \quad f(r) = \frac{r^2}{L^2}\left(  1 + \frac{L^2\mu^2r_h^2 }{r^4} -\frac{L^4\beta^2 }{2r^2}  -\frac{r_h^3}{r^3} \left( 1 + \frac{L^2\mu^2 }{r_h^2} -\frac{L^4\beta^2 }{2r_h^2}  \right)   \right)  \,, \\
A&= \mu \left(  1- \frac{r_h}{r}   \right)\mathrm{d} t    \,, \qquad \psi_1  = \beta x \,, \qquad \psi_2  = \beta y\,,
\end{split}
\end{equation}
where $\mu$ is the chemical potential and $\beta$ is the strength of momentum relaxation. $r_{h}$ is the black hole horizon, which can be expressed, using the Hawking temperature $T=f'(r_{h})/4\pi$, as
\begin{equation} \label{rh1}
r_h = \frac{L^2}{6}\left(4\pi T+\sqrt{6\beta^2+16\pi^2 T^2+12L^{-2}\mu^2}\right)\,.
\end{equation}

As in the Gubser-Rocha-linear axion model, we investigate the spectral functions at fixed chemical potential: ($T/\mu, \beta/\mu$). In this background we consider the fermion action in the same way as in section \ref{method123}. 

\begin{figure}[]
\centering
     {\includegraphics[width=13cm]{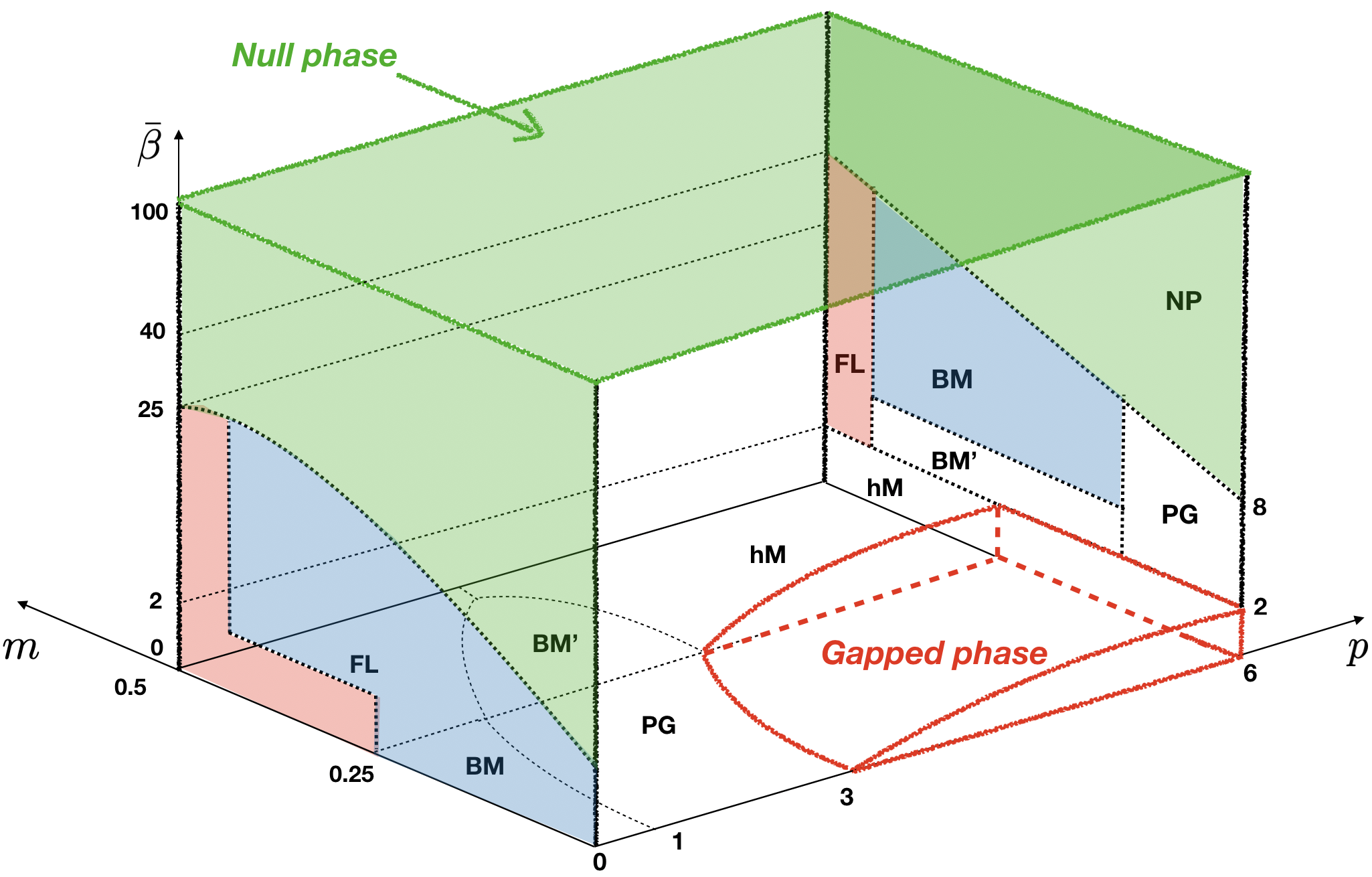} \label{}}
 \caption{Phase diagram in ($m, p, \bar{\beta}$) space. }\label{CompletePhaseD222}
\end{figure}

\begin{figure}[]
\centering
     \subfigure[$p=0$]
     {\includegraphics[width=4.82cm]{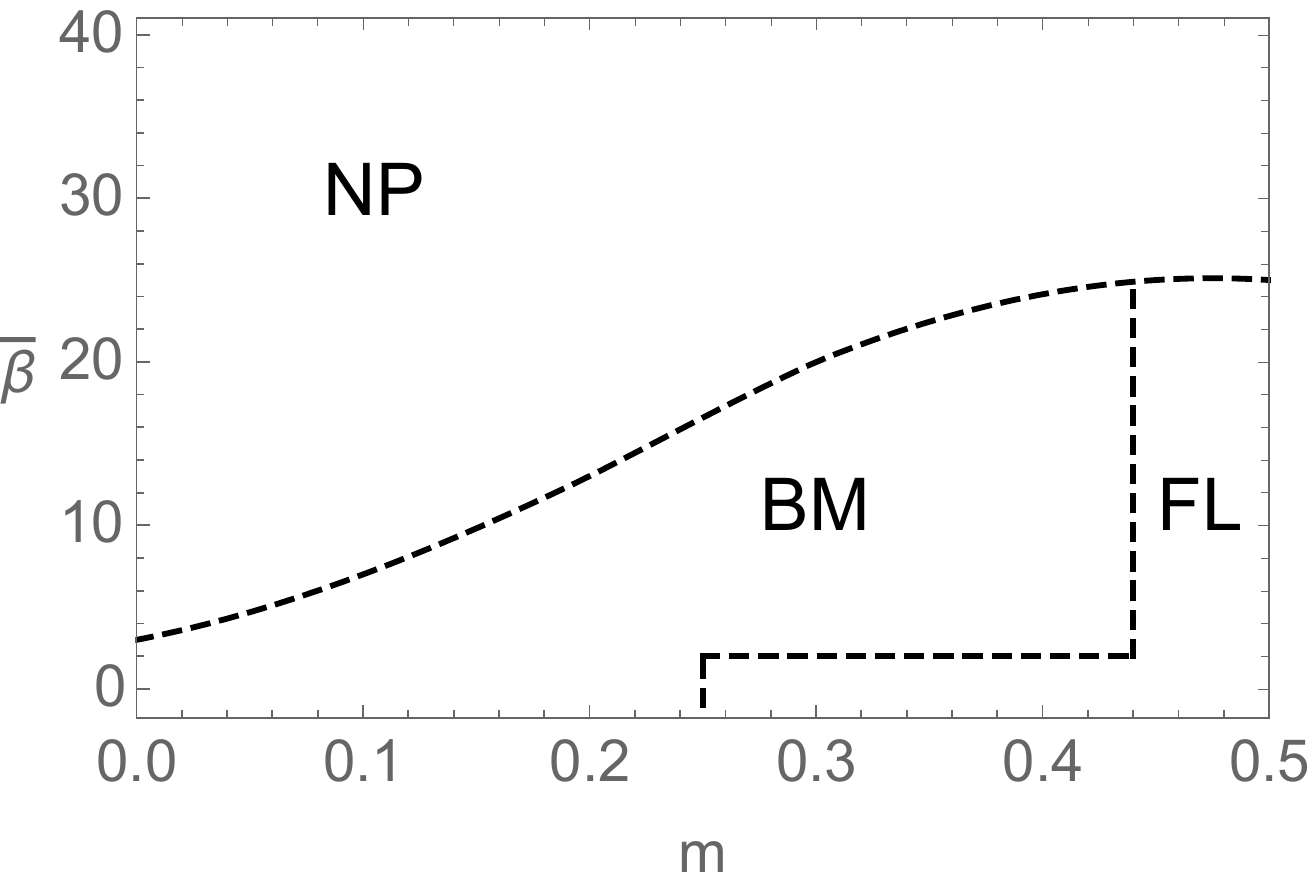} \label{}}
     \subfigure[$p=3$]
     {\includegraphics[width=4.82cm]{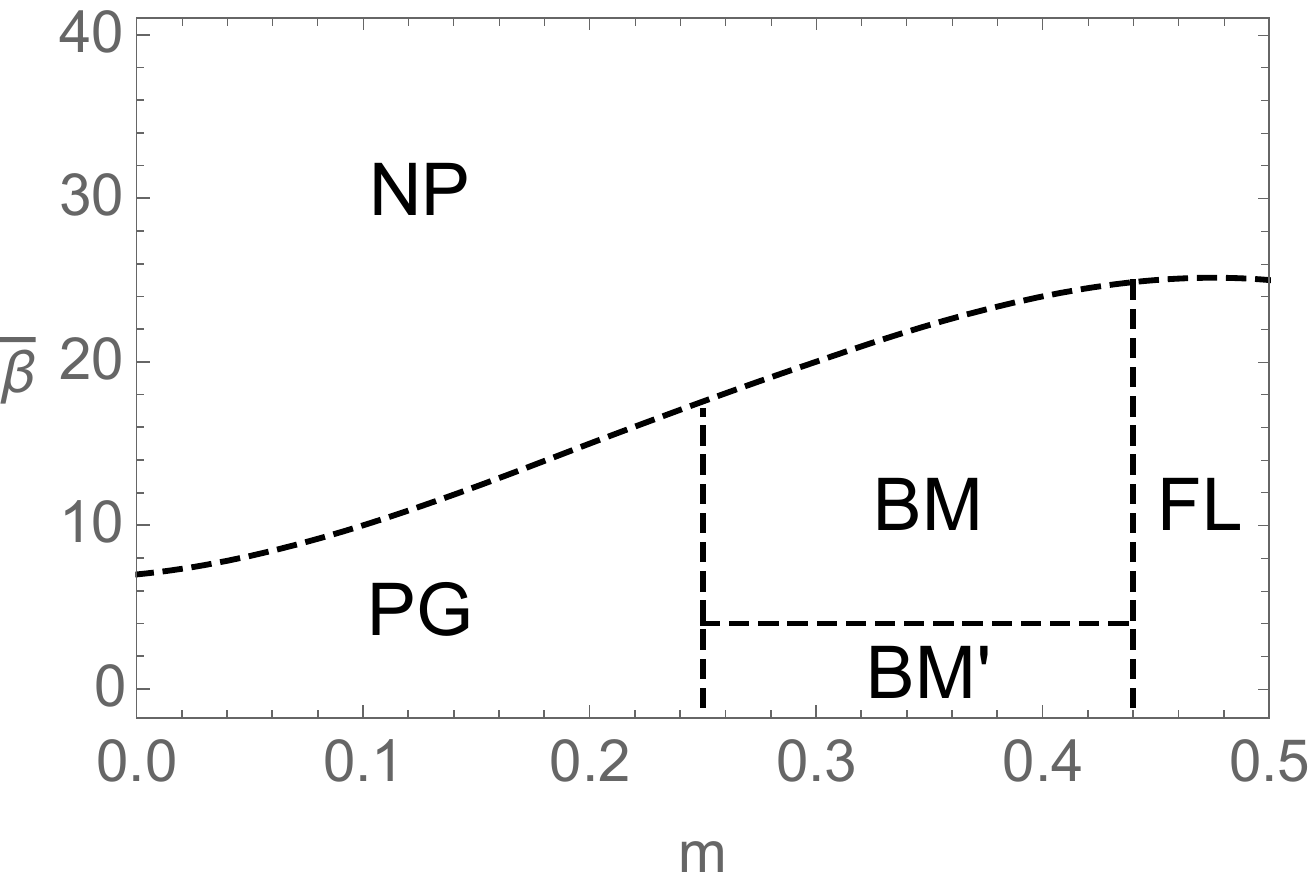} \label{}}
     \subfigure[$p=6$]
     {\includegraphics[width=4.82cm]{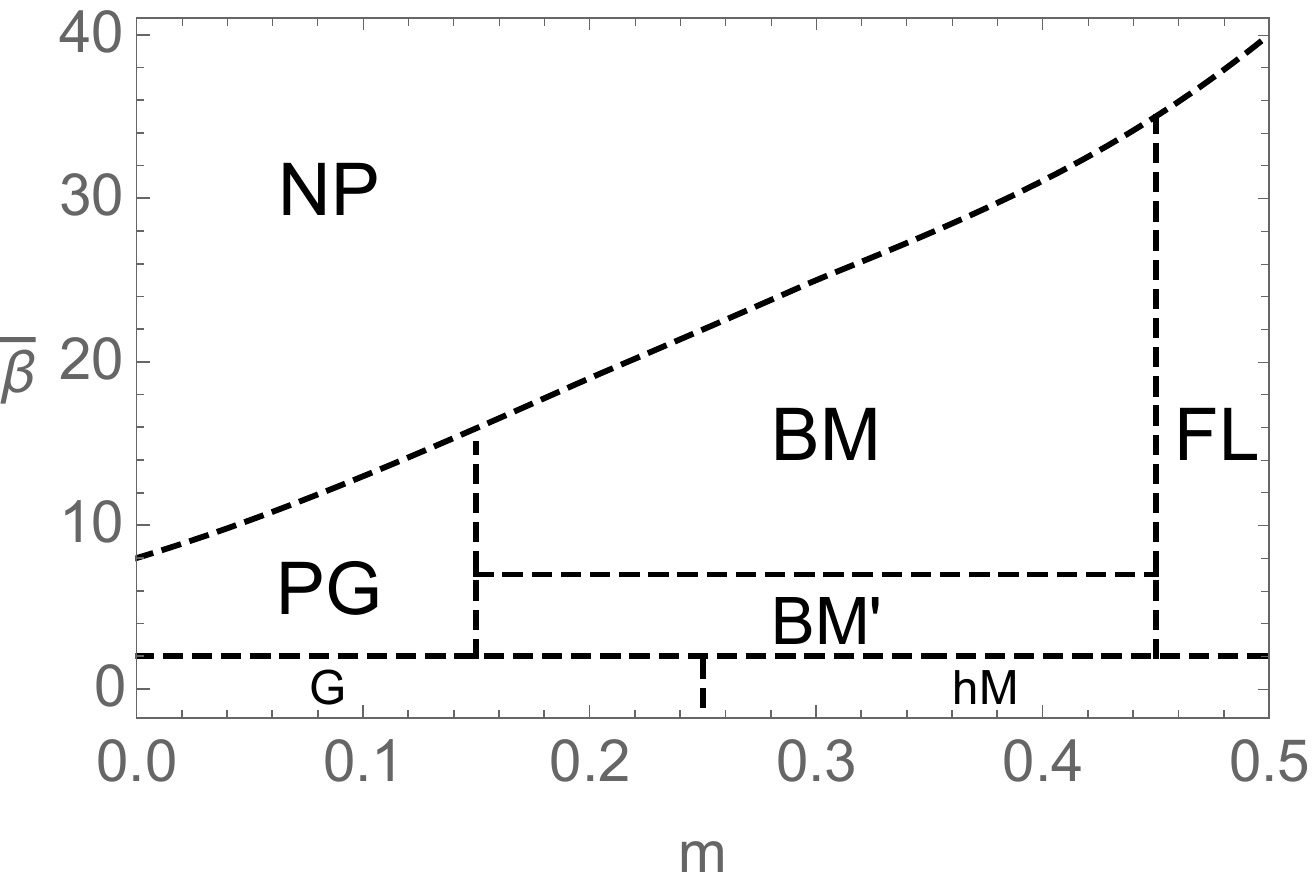} \label{}}
 \caption{Phase diagram in ($m$, $\bar{\beta}$) space. }\label{FIG92}
\end{figure}

\begin{figure}[]
\centering
     {\includegraphics[width=15.3cm]{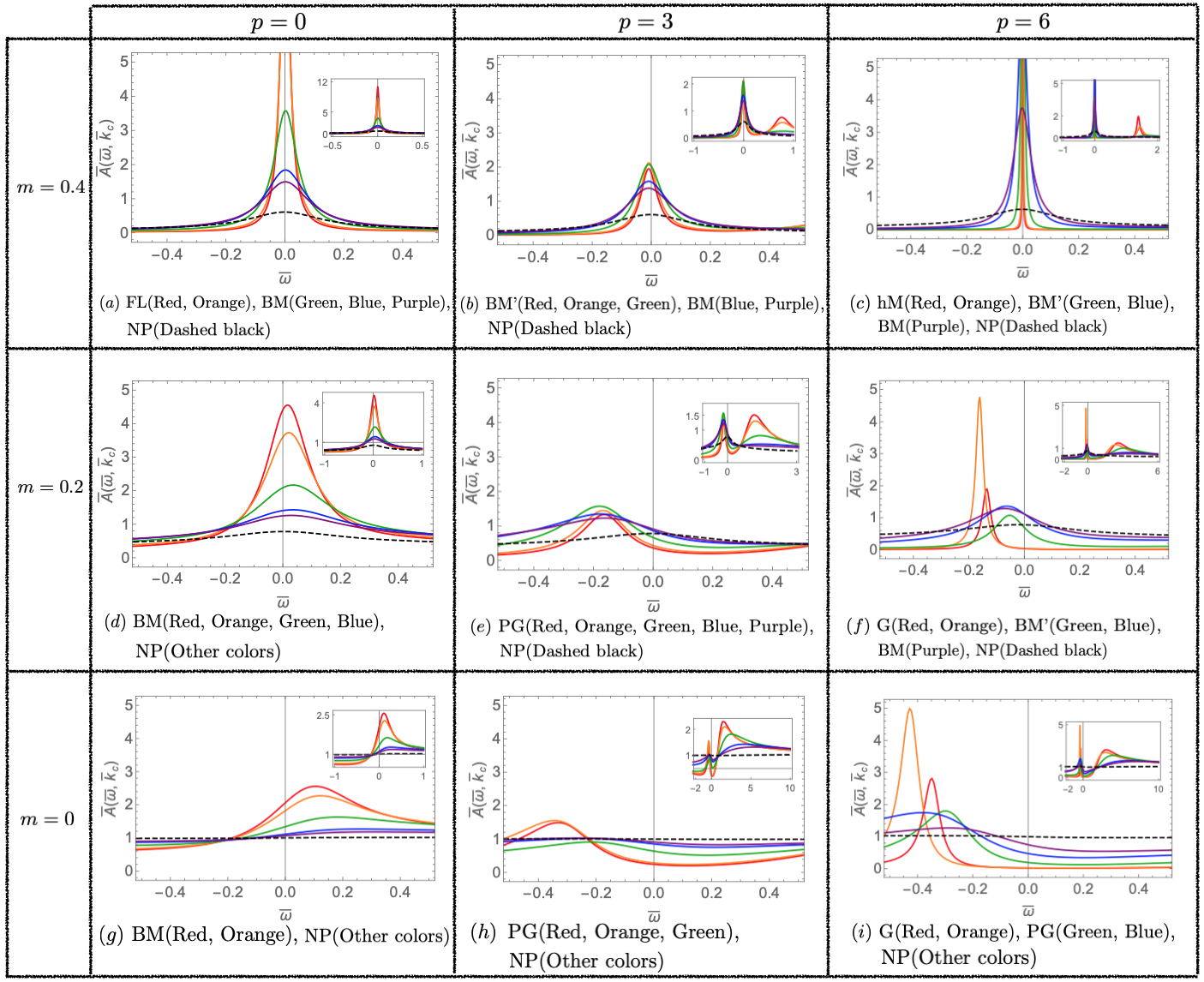} \label{}}
\caption{Spectral functions at finite momentum relaxation. Different color refers to different momentum relaxation: $\bar{\beta}$ = 0, 1, 3, 7, 10, 100 (red, orange, green, blue, purple, dashed black). }\label{LAMPD2} 
\end{figure}

For a parallel analysis with the previous section \ref{ADM123}, we display the spectral functions for $m=(0, 0.2, 0.4)$ and $p=(0, 3, 6)$ in Fig. \ref{LAMPD2}.
As momentum relaxation becomes strong, phases are changed as
\begin{align}\label{CASES2}
\begin{split}
&\text{(a): FL} \rightarrow \text{BM} \rightarrow \text{NP} \,, \,\,\,\,\,  \text{(b): BM'} \rightarrow \text{BM} \rightarrow \text{NP} \,, \,\,\,\, \text{(c): hM} \rightarrow  \text{BM'} \rightarrow \text{BM} \rightarrow \text{NP} \,, \\
&\text{(d): BM} \rightarrow \text{NP} \,, \qquad\quad\,\,\,\,  \text{(e): PG}  \rightarrow \text{NP} \,, \quad\qquad\,\,\,\,\,\,\,\, \text{(f): G} \rightarrow  \text{BM'} \rightarrow \text{BM} \rightarrow  \text{NP} \,, \\
&\text{(g): BM} \rightarrow  \text{NP}  \,, \qquad\quad\,\,\,\,  \text{(h): PG} \rightarrow \text{NP} \,, \quad\qquad\,\,\,\,\,\,\,\, \text{(i): G} \rightarrow \text{PG} \rightarrow \text{NP} \,. \\
\end{split}
\end{align}
Comparing \eqref{CASES2} with \eqref{CASES}, we conclude that the linear axion model shows the qualitatively same behavior as the Gubser-Rocha-linear axion model when momentum relaxation becomes stronger: i) the spectral function is suppressed; ii) all fermionic phases become NP.
Similarly to Fig. \ref{CompletePhaseD22}, we can also identify all phases in three dimensional space $(m,p,\bar{\beta})$.  See Fig.~\ref{CompletePhaseD222} and some of its cross-sections in Fig. \ref{FIG92}.
%


%

Let us finish this section by making comments on three common features both in the Gubser-Rocha-linear axion model and the linear axion model. 

\paragraph{Mass effect: sharpening spectral functions}
For a given dipole coupling and momentum relaxation, the spectral functions of both models tend to be sharper as the fermion mass increases. This seems related to the fact that when the fermion mass is $m=0.5$, which is the maximum mass allowed by the unitary bound, 
the conformal dimension of the dual operator becomes unity, which is of a free fermion. If $p=\beta=0$, the spectral function becomes delta-function like. As $m$ decreases, the spectral function becomes broader. For example, see the red curves in the left column of Fig. \ref{m0FIG} and Fig. \ref{LAMPD2}. It is a general feature also for $p \ne 0$ or $\beta \ne 0$ in gapless phase.\footnote{In pseudogap and gapped phase, the definition of spectral function is ambiguous because there is no fermi momentum as explained in sec. \ref{sec31}.} 

\begin{figure}[]
\centering
     {\includegraphics[width=15.2cm]{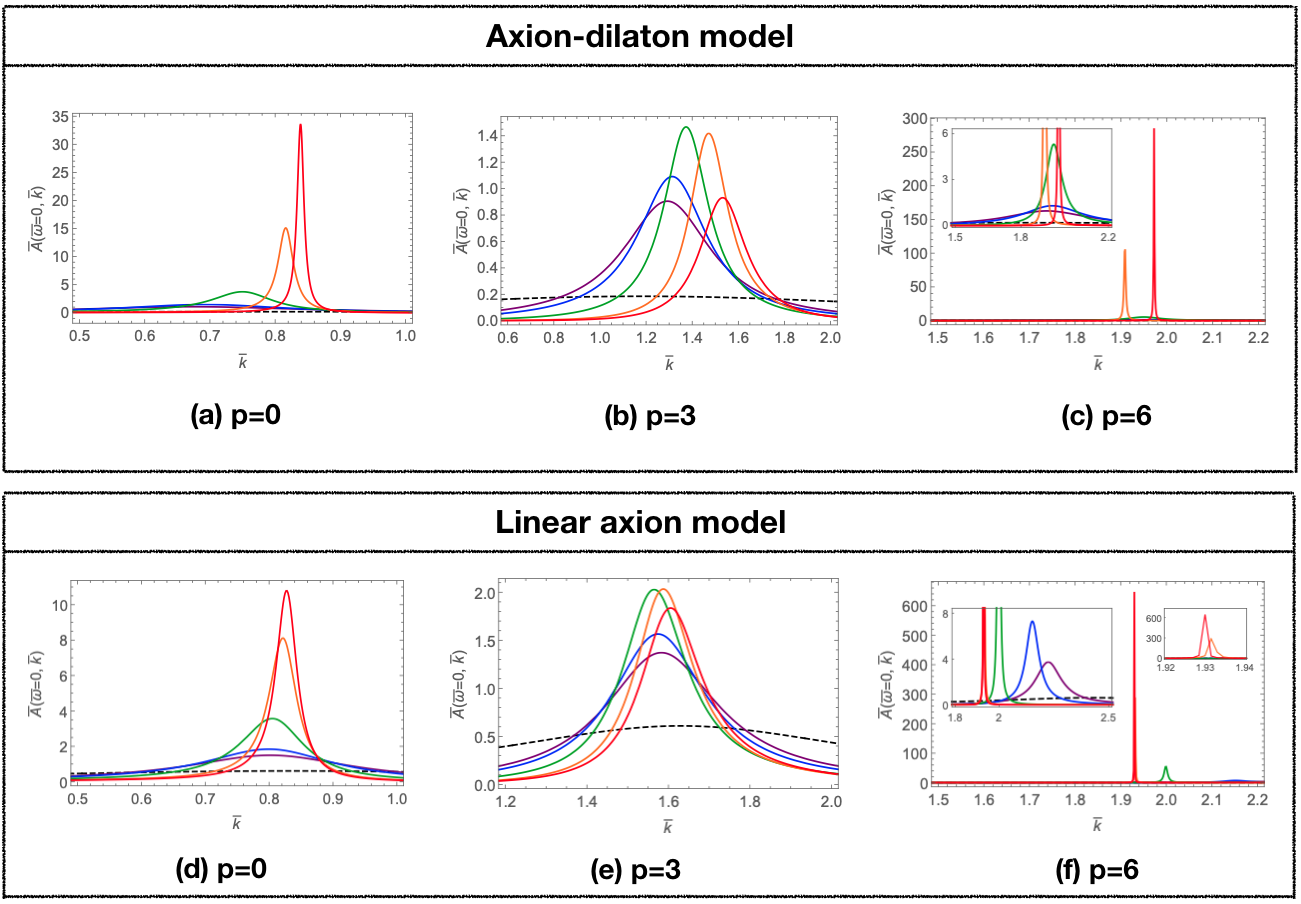} \label{}}
 \caption{Spectral function of both models at $\bar{\omega}=0$ for $m=0.4$. Different colors represent various momentum relaxation: $\bar{\beta} = 0\,,1\,,3\,,7\,,10\,,100$ (red, orange, green, blue, purple, black) }\label{LAMPD1}
\end{figure}

\paragraph{Momentum relaxation effect: broadening spectral functions}
In general the spectral function becomes more suppressed and broader as momentum relaxation increases. However, in the intermediate $p$ regime (approximately $1<p<5$), the broadening effect of $\beta$ is not manifest due to finite $p$ and $m$ effect. 
It can be seen in Fig. \ref{m0FIG} and Fig. \ref{LAMPD2} but, to display it more clearly, we make plots of spectral functions as function of $\bar{k}$ at $\bar{\omega}=0$ at $m=0.4$ in Fig. \ref{LAMPD1}.
Especially, (b)(e) of Fig. \ref{LAMPD1} show different behaviors of spectral function with increasing $\bar{\beta}$. First, it increases up to certain value of $\bar{\beta}$(e.g., Green line), and then decreases. 
%
%

\paragraph{Dipole coupling effect: new peak or gap-generation}
As the dipole coupling increases, a new peak at finite frequency is generated. If mass is big (close to $m=0.5$) the original peak at $\bar{\omega}=0$ still remains (so metalic) while if mass is small (close to $m=0$) the original peak at $\bar{\omega}=0$ is shifted to the negative values (so pseudogap or gap).

\section{Conclusion} \label{}
In this paper, we have investigated the holographic spectral function with momentum relaxation in two holographic models: the Gubser-Rocha-linear axion model and the linear axion model. 
We have computed holographic spectral functions by choosing several values of three parameters:
i) fermion mass ($m$), ii) dipole coupling ($p$), iii) the strength of momentum relaxation ($\beta$). Depending on the shape of spectral functions, we classify the phases as: fermi liquid like (FL), bad metal prime (BM'), bad metal (BM), pseudogap (PG), gapped (G) as shown in Fig. \ref{Typical}. By this classification, we may construct a phase diagram in ($m,p,\beta$) space. 

We find that both models show similar phase diagram in $(m,p,\beta)$ space. See Fig. \ref{CompletePhaseD22} and Fig. \ref{CompletePhaseD222}. In both models, a larger fermion mass (close to $m=0.5$) makes the spectral function sharper while a larger momentum relaxation makes the spectral function broader. A dipole coupling generates a new pick at finite frequency and supports psedogap or gapped phase for small mass.  More common effects of the parameters are summarized in Table. \ref{table1} and Table. \ref{table2}.

%
\begin{table}[]
\begin{tabular}{| p{7cm} | p{7.1cm} | p{7.0cm} |}
\hline
            \qquad\qquad\qquad\quad  Increasing $m$     & \qquad\qquad\qquad\quad    Increasing $p$ \\ 
 \hline
 \hline
- Gapped phase becomes gapless phase.  &  - New gapless phases appears for large $m$. \\ 
- Larger $p$ is required to open the gap.  &  - Gapped phase is generated for small $m$.  \\ 
 \hline
\end{tabular}
\caption{ Spectral function \textit{without} momentum relaxation ($\beta=0$).}
\label{table1}
\end{table}
\begin{table}[]
\begin{tabular}{| p{14.5cm} | p{14.5cm} | p{14.5cm} |}
\hline
           \qquad\qquad\qquad\qquad\qquad\qquad\qquad\quad\quad  Increasing $\beta$     \\ 
 \hline
 \hline
- The spectral function is suppressed in the whole range of ($m, p$).   \\
- The various phases become featureless (NP) at strong momentum relaxation.      \\ 
 \hline
\end{tabular}
\caption{Spectral function \textit{with} momentum relaxation $\beta \ne 0$.}
\label{table2}
\end{table}

\begin{figure}[]
\centering
     {\includegraphics[width=8cm]{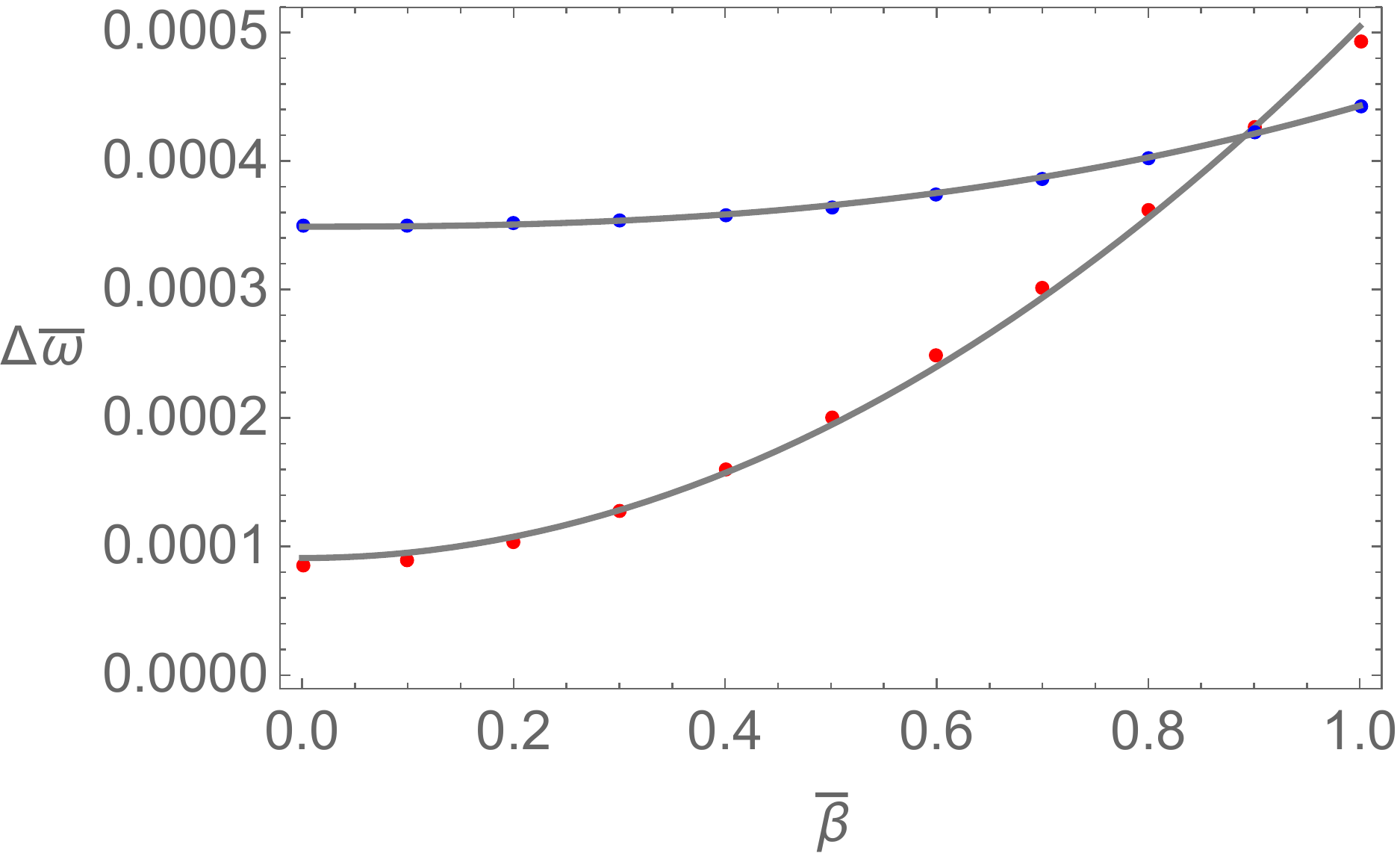}}
 \caption{The width of the spectral functions vs momentum relaxation. The red dots ($\sim 0.0001+0.0004 \bar{\beta}^2$) are for the Gubser-Rocha-linear axion model.  The blue dots ($\sim 0.00035+0.000095 \bar{\beta}^{2.5}$) are for the linear axion model.}\label{width}
\end{figure}

It was noted that if $m=0.5$, the conformal dimension of the dual operator becomes 1, which is of a free fermion and the spectral function becomes delta-function like. As momentum relaxation is increased the spectral functions become broader.  For example, we show the change of the width as functions of momentum relaxation  in Fig. \ref{width}, where we choose $m=0.498$ for numerical stability instead of $m=0.5$.
The red dots are for the Gubser-Rocha-linear axion model and the gray fitting curve is: $\sim 0.0001+0.0004 \, \bar{\beta}^2$.  The blue dots are for the linear axion model and the gray fitting curve is: $\sim 0.00035+0.000095 \, \bar{\beta}^{2.5}$. In weak impurity scattering process in the Fermi liquid theory, the width of the spectral function can be related to the inverse of relaxation time $1/\tau$ which is proportional to the impurity density, $n_{imp}$. Even though the Fermi-liquid theory is not applicable, just to have an intuition, we may compare our results with the Fermi liquid theory. Then, we may say the effective impurity density is $n_{imp} \sim \beta^2$ for the Gubser-Rocha-linear axion model and $n_{imp} \sim \beta^{5/2}$ for the linear axion model.

It will be interesting to see if there is any relation between holographic spectral function and holographic conductivity, because they are intimately related in field theory sides. For this comparison let us briefly review DC conductivities in two models we have considered. 
%

For the Gubser-Rocha-linear axion model and the linear axion model, DC electric conductivities are given by~\cite{Jeong:2018tua,Andrade:2013gsa}
\begin{align}
&\sigma_{DC} =  \sqrt{1+\frac{1}{\tilde{r}_{h}}}\left( 1 + \frac{1}{\bar{\beta}^2} \right) \,, \,\,\quad \text{(Gubser-Rocha-linear axion model)}  \label{conduct1}\\
&\sigma_{DC} =   1 + \frac{1}{\bar{\beta}^2} \,,                              \quad\qquad\qquad\qquad \text{(Linear axion model)}  \label{conduct2}
\end{align}
where $\tilde{r}_{h} :=  r_{h}/Q$ and $\bb := \beta/\mu$. In particular, for strong momentum relaxation ($\bar{\beta}\gg1$), 
the formula becomes

\begin{figure}[]
\centering
     {\includegraphics[width=8cm]{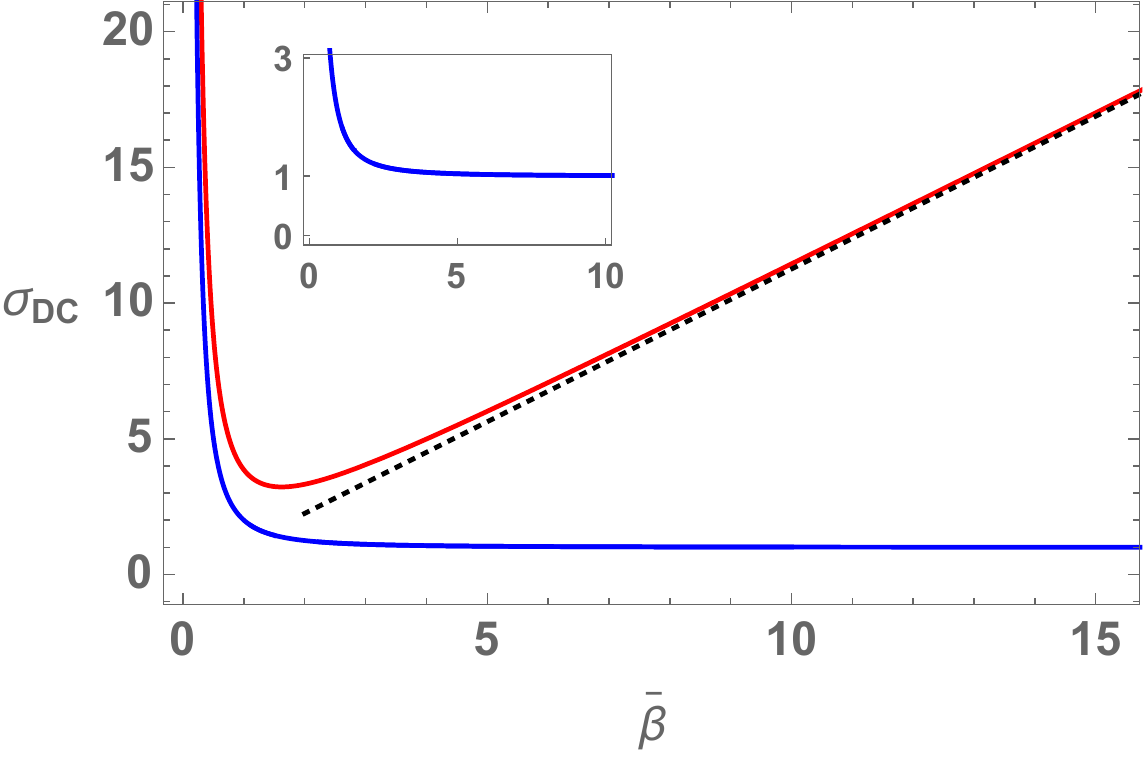} \label{}}
 \caption{The DC electric conductivity for the Gubser-Rocha-linear axion model (red line) and the linear axion model (blue line). The red and blue solid lines represent  \eqref{conduct1} and \eqref{conduct2}, respectively. The dashed black line is \eqref{GubForm}, the Gubser-Rocha-linear axion model in $\bar{\beta}(=\beta/\mu)\gg1$ limit. The inset figure shows that $\sigma_{DC}$ of the linear axion model approaches to 1 with increasing $\bar{\beta}$.}\label{CONFIG}
\end{figure}
%
%
\begin{align}
&\sigma_{DC} \sim  \sqrt{\frac{1}{\tilde{r}_{h}}} \sim \frac{\bb}{2\sqrt{2}\pi \bT}  \,, \,\, \qquad \text{(Gubser-Rocha-linear axion model)}  \label{GubForm}\\
&\sigma_{DC} \sim  1  \,, \,\,\,\,\,\qquad\qquad\qquad\qquad \text{(Linear axion model)} \label{LAMForm}
\end{align}
since
\begin{align}
&\tilde{r}_{h} \sim \frac{8\pi^2 \bT^2}{\bb^2} \,, \qquad \qquad \quad (\bb \gg 1) \, \label{tQ3}
\end{align}
for the Gubser-Rocha-linear axion model~\cite{Jeong:2018tua}.
At a given temperature $\bar{T}=1/10$, the DC electric conductivities for two models are plotted in Fig. \ref{CONFIG}. The dashed black line in the figure is \eqref{GubForm}.

Note that, as momentum relaxation increases, the DC conductivity increases in the Gubser-Rocha-linear axion model while decreases and saturates to unity in  the linear axion model.  
Because of this difference in conductivity we may expect that fermionic spectral functions of two models may behave differently as we increase momentum relaxation. However, our results show that the effect of momentum relaxation is the same for both models:  it simply suppresses spectral function. This universal\footnote{It has been shown that the spectral functions are suppressed at finite momentum relaxation in limited range of parameters ($m,p$) also in different models~\cite{Bagrov:2016cnr, Ling:2014bda, Cremonini:2018xgj, Cremonini:2019fzz}.} feature may be due to the fact that holographic fermion does not back-react to geometry while gauge fields used in computing conductivity back-react to geometry.

%

We also may expect that the phase diagram of two models can be different at zero momentum relaxation at zero temperature because their IR geometries are different. It turns out that at low temperature, $T/\mu=0.1$, the phase diagrams of both models are qualitatively the same. See the bottom of Fig. \ref{CompletePhaseD22} and Fig. \ref{CompletePhaseD222}. However, it is possible that they become different in the zero temperature limit.

%

%
%

It will be interesting to find how our holographic results can be related to experimental results in condensed matter systems such as correlated electron systems with disorder~\cite{Byczuk:2005aa}, disorder induced polaron~\cite{Emin:1994aa} and impurity-induced states in unconventional superconductors~\cite{Balatsky:2006aa}.
Moreover, it will be interesting to study how our results can be related to other holographic quantities such as quasinormal modes of the fermionic Green's function~\cite{Bagrov:2016cnr} and AC conductivity~\cite{Ammon:2019wci, Baggioli:2019aqf, Donos:2019tmo, Amoretti:2018tzw}. 
In this paper, we consider a dipole coupling as an interaction term, 
but it is also interesting to study the momentum relaxation effect on other types of interaction such as a Majorana coupling~\cite{Faulkner:2009am}.
Another future direction of this paper is to investigate the temperature dependence of spectral functions in $(m, p, \beta)$ space\footnote{Ref. \cite{Fang:2015dia} studied the temperature dependence of spectral function in $(m = 0, p = 0, \beta)$ space and found the phase transition between Fermi liquid phase and non-Fermi liquid phase.}.
We leave these subjects as future works.

\acknowledgments

This work was supported by Basic Science Research Program through the National Research Foundation of Korea(NRF) funded by the Ministry of Science, ICT $\&$ Future Planning(NRF- 2017R1A2B4004810) and GIST Research Institute(GRI) grant funded by the GIST in 2019. 
YS was supported by Basic Science Research Program through NRF grant No. NRF-2019R1I1A1A01057998.
SJS is supported by Mid-career Researcher Program through the National Research Foundation of Korea grant No. NRF-2016R1A2B3007
687.
S-Y Wu was supported by MOST 105-2112-M-009-001-MY3. We also would like to thank the APCTP(Asia-Pacific Center for Theoretical Physics) focus program, ``Quantum Matter from the Entanglement and Holography" in Pohang, Korea for the hospitality during our visit, where part of this work was done.



\providecommand{\href}[2]{#2}\begingroup\raggedright\endgroup

\end{document}